\documentclass[12pt,a4paper,english]{article}
\usepackage[T1]{fontenc}
\usepackage[latin1]{inputenc}

\makeatletter

\providecommand{\tabularnewline}{\\}

\usepackage{subfigure}
\usepackage{graphicx}

\usepackage[T1]{fontenc}

\newcommand{\beq}{\begin{equation}}
\newcommand{\eeq}{\end{equation}}

\newcommand{\ZZ}{\mathbb{Z}}
\newcommand{\RR}{\mathbb{R}}

\newcommand{\hlf}{\frac{1}{2}}

\newcommand{\ket}[1]{|\, #1 \rangle}

\usepackage{amsfonts}
\usepackage{latexsym}

\usepackage{babel}
\makeatother
\begin{document}
\title{N=1 Supersymmetric Boundary Bootstrap }

\date{21st August 2003}
\author{G\'{a}bor Zsolt T\'{o}th}

\maketitle
\begin{center}\textsl{Institute for Theoretical Physics, }\\
\textsl{E\"{o}tv\"{o}s University, Budapest, Hungary}\\
\textsl{H-1117 Budapest, P\'{a}zm\'{a}ny P\'{e}ter S\'{e}t\'{a}ny
1/A}\end{center}

\begin{center}\texttt{tgabor@afavant.elte.hu}\end{center}

\begin{abstract}
We investigate the boundary bootstrap programme for finding exact
reflection matrices of integrable boundary quantum field theories
with N=1 boundary supersymmetry. The bulk S-matrix and the reflection
matrix are assumed to take the form $\hat{S}=\tilde{S}S$, $\hat{R}=\tilde{R}R$,
where $\tilde{S}$ and $\tilde{R}$ are the S-matrix and reflection
matrix of some integrable non-supersymmetric boundary theory that
is assumed to be known, and $S$ and $R$ describe the mixing of supersymmetric
indices. Under the assumption that the bulk particles transform in
the kink and boson/fermion representations and the ground state is
a singlet we present rules by which the supersymmetry representations
and reflection factors for excited boundary bound states can be determined.
We apply these rules to the boundary sine-Gordon model, to the boundary
$a_{2}^{(1)}$ and $a_{4}^{(1)}$ affine Toda field theories, to the
boundary sinh-Gordon model and to the free particle. 
\end{abstract}
PACS codes: 11.10.Kk, 11.55.Ds, 11.30.Pb\\
Keywords: supersymmetry, integrable quantum field theory, boundary
scattering, bound state, bootstrap, sine-Gordon model, sinh-Gordon
model, affine Toda theory

\section{Introduction}

Integrable quantum field theories (QFTs) in two dimensions have the
remarkable property that their S-matrix factorizes and the bootstrap
programme becomes manageable. The particle spectrum and S-matrix of
several integrable QFTs have been found by applying the axioms of
general S-matrix theory and factorized scattering theory. Studying
supersymmetric field theories it is an interesting problem to determine
the S-matrix and particle spectrum of integrable supersymmetrized
QFTs.

A formalism for constructing $N=1$ supersymmetric factorizable scattering
theory is developed in \cite{key-3,key-1,key-15,key-22,key-23}. The
particles of the non-supersymmetric theory are assumed to become supermultiplets
$\ket{\xi_{i}(\theta)}\otimes\ket{A_{i}(\theta)}$, where $\ket{\xi_{i}(\theta)}$
are the one-particle states of the non-supersymmetric theory, and
it is also assumed that all the particles of the supersymmetric theory
are obtained in this way. The index $i$ specifies (not necessarily
all) Poincare-invariant conserved quantum numbers of the non-supersymmetric
theory and $A_{i}$ specifies the supersymmetric quantum numbers.
Those conserved quantum numbers which are specified by $i$ must not
change their value under charge conjugation, and supersymmetry (except
for boosts) acts trivially on the $\ket{\xi_{i}(\theta)}$ part of
the states. As the simplest Ansatz the product form $\hat{S}_{\xi_{i}A_{i};\xi_{j}A_{j}}^{\xi_{j}'A_{j}';\xi_{i}'A_{i}'}(\theta_{1}-\theta_{2})=\tilde{S}_{\xi_{i}\xi_{j}}^{\xi_{j}'\xi_{i}'}(\theta_{1}-\theta_{2})S_{A_{i}A_{j}}^{A_{j}'A'_{i}}(\theta_{1}-\theta_{2})$
was proposed for the two-particle S-matrix $\hat{S}$, where $\tilde{S}_{\xi_{i}\xi_{j}}^{\xi_{j}'\xi_{i}'}(\theta)$
is the two-particle S-matrix of the non-supersymmetric theory subject
to supersymmetrization, and the supersymmetric factor $S_{A_{i}A_{j}}^{A_{j}'A_{i}'}(\theta)$
describes the mixing of supersymmetric indices (see also the first
paragraph of section 2 for conventions). The requirement that the
particle spectrum can be obtained as described above is consistent
with the Ansatz for the S-matrix if {}``minimal'' supersymmetric
S-matrix factors having no poles and overall zeroes on the imaginary
axis in the physical strip are used. We restrict ourselves to such
S-matrix factors (see section \ref{sec BShG} for an exception). As
the notation indicates, it is required that the SUSY representations
associated to the particles depend on conserved Poincare- and charge
conjugation invariant quantum numbers only, what guarantees that the
complete S-matrix (including amplitudes of multi-particle scatterings)
of the supersymmetric theory also takes a product form. The complete
two-particle S-matrix $\hat{S}$ and so the supersymmetric factor
$S$ commute with the action of supersymmetry. The Yang-Baxter, fusing,
and bootstrap equations, the crossing equation, the unitarity and
analyticity conditions are satisfied by $\hat{S}$ and take a factorized
form. Since $\tilde{S}$ satisfies them, the supersymmetric factor
$S$ must also satisfy them separately. The fusing angles occurring
in the fusing and bootstrap equations for $S$ are determined by the
non-supersymmetric theory, and since the bound state poles are already
present in $\tilde{S}$, the supersymmetric factor $S$ satisfies
the bootstrap equations passively, i.e. $S$ does not have poles at
the fusing rapidities.

An essential step in this construction is the choice of SUSY representations
in which the particles transform. Having made this choice the supersymmetric
factor $S$ of the two-particle S-matrix is obtained by solving the
supersymmetry condition, the Yang-Baxter and crossing equations, the
unitarity, analyticity and minimality conditions. These equations
depend on the conserved quantum numbers, so they impose restrictions
on the choice of representations. Further highly nontrivial restrictions
come from the bootstrap and especially from the fusing equations for
$S$, which contain the fusing data of the non-supersymmetric model.
Let $a_{i}+b_{j}\rightarrow c_{k}$ be a fusion process of the non-supersymmetric
theory with fusing rapidity $iu_{a_{i}b_{j}}^{c_{k}}$, and assume
that the representations in which $a_{i}$, $b_{j}$, and $c_{k}$
transform are $D_{i}$, $D_{j}$ and $D_{k}$; in this case the fusing
equation is equivalent to the statement that the supersymmetric two-particle
S-matrix block at rapidity $iu_{a_{i}b_{j}}^{c_{k}}$ is an equivariant
projection from the product of $D_{i}$ and $D_{j}$ on $D_{k}$.
This can be very restrictive for the value of $u_{a_{i}b_{j}}^{c_{k}}$,
as in the most common case, for example, when $D_{i}$, $D_{j}$,
$D_{k}$ are boson/fermion representations \cite{key-1,key-3}. We
remark that the fusing equation can be forced to hold in certain cases
by factoring out unwanted states from the Hilbert-space.

The previous considerations indicate that it is nontrivial that the
construction above can be applied to supersymmetrize a given factorized
scattering theory. The knowledge of the Lagrangian density or other
data of the (supersymmetric) theory often suggests preferred SUSY
representations for the particle multiplets. If the possible representations
in which the particles can transform are fixed, then by solving the
axioms above one can derive necessary and sufficient conditions that
have to be satisfied by the particle spectrum and fusing rules of
a non-supersymmetric theory which is supersymmetrized using these
representations. These conditions have been obtained in the case when
the possible representations are the kink and the boson/fermion representation,
and several factorizable scattering theories ---the $a_{n-1}^{(1)}$,
$d_{n}^{(1)},$ $(c_{n}^{(1)},d_{n+1}^{(2)})$ and $(b_{n}^{(1)},a_{2n-1}^{(2)})$
affine Toda theories, the $SU(2)$ principal chiral model, the sine-Gordon
model, the supersymmetric $O(2n)$ sigma model \cite{key-1}, and
the multicomponent Yang-Lee (or FKM) minimal models \cite{key-4}---
have been found to satisfy these conditions, i.e. these theories have
been supersymmetrized by applying the construction described above.
The Lagrangian field theories underlying these supersymmetric scattering
theories are not all known (if they exist).

It is a natural step after the study of bulk factorized scattering
theories to study them in the presence of a boundary. If the boundary
conditions and interactions preserve integrability, then the bootstrap
programme remains manageable. A set of boundary bound states are added
to the bulk spectrum, and scattering processes involving boundary
states are described by reflections amplitudes which can be written
as products of one-particle reflection amplitudes and two-particle
S-matrices. The boundary versions of the unitarity condition, crossing
equation, Yang-Baxter, fusing, and bootstrap equations can be introduced
\cite{key-17}. The Ansatz above for constructing supersymmetric scattering
theories can also be extended to the situation when an integrable
boundary is present. A nontrivial problem of this extension that we
discuss in this paper is the definition of the concept of supersymmetry
in the presence of boundary. Results in the classical Lagrangian field
theory framework \cite{key-8} indicate clearly that such a concept
is possible. Our construction is based on the results in \cite{key-13,key-14,key-2,key-4,key-8,key-9}
and it can be regarded as an application of the construction described
in \cite{key-13,key-14} for quantum group symmetry.

Let us assume that a factorizable scattering theory with boundary
is given and the supersymmetric version of the bulk part of this theory
is also constructed using the Ansatz above, the bulk particles transforming
either in the kink or in the boson/fermion representation. We study
in this paper the problem of completing the supersymmetrization in
this case using the boundary version of the Ansatz above. The first
step is to choose a representation of the boundary supersymmetry algebra
for the ground state. Next the minimal supersymmetric factors of the
ground state one-particle reflection amplitudes are to be determined
using the boundary Yang-Baxter, unitarity and crossing equations and
the supersymmetry condition for these factors. Finally the boundary
bootstrap and fusing equations for supersymmetric factors can be used
to obtain the representations in which the excited boundary bound
states transform together with the supersymmetric factors of the one-particle
reflection amplitudes on these states. We remark that if there were
preferred representations for the excited boundary states, then the
boundary fusing equation would constrain the boundary fusing angles.
The first and especially the second steps have been considered in
the literature \cite{key-10,key-11,key-4,key-5,key-9,key-2}, whereas
there are fewer results concerning the last step \cite{key-2}. There
are few known solutions to the boundary bootstrap programme for non-supersymmetric
theories either.

For the ground state we take the singlet representations with RSOS
label $\hlf$, this being the simplest case and also because we expect
\cite{key-2,key-10} that other cases can be obtained by boundary
bootstrap from this one. The general minimal supersymmetric one-particle
ground state reflection factors have been determined for this case
in \cite{key-10,key-11,key-5} but without imposing the supersymmetry
condition. This condition is imposed on the kink reflection amplitude
in \cite{key-2}. In the present paper we consider the boson/fermion
ground state reflection factors in some detail. We derive these reflection
factors by imposing the supersymmetry condition and solving the boundary
Yang-Baxter equation afterwards.

As the main result of the paper we determine the representations and
supersymmetric one-particle reflection factors for excited boundary
bound states using the bootstrap and fusing equations, for arbitrary
boundary fusing rules. This is a nontrivial problem because the one-particle
reflection amplitudes and so the fusing tensors can be degenerate
at particular rapidities. Although these rapidities are special, they
usually play an essential role in boundary theories. We also present
the conditions on the fusing angles resulting from the requirement
of minimality. This requirement is nontrivial even if the ground state
reflection factors are minimal, and restricts the boundary fusing
angles. Our work is motivated by the results of \cite{key-1} for
bulk supersymmetric bootstrap, and by \cite{key-2}, where the scattering
theory of the boundary sine-Gordon model (BSG) is supersymmetrized.
In the BSG theory the whole boundary spectrum can be generated by
kinks, and in this case it is not difficult to determine the representations
and supersymmetric reflection factors for the boundary states. However,
there are other ways to generate the boundary states, and it is nontrivial
---though plausible--- that the same representations and reflection
factors are obtained in each way. The correctness of this proposition
for the BSG model was conjectured and partially verified in \cite{key-2}.
Using our results we can complete the verification and confirm that
the proposition is correct.

Only few integrable boundary scattering theories are known explicitely.
Beyond the BSG model non-trivial scattering theories for boundary
$a_{2}^{(1)}$ and $a_{4}^{(1)}$ affine Toda field theories \cite{key-7},
for the free boson on the half line, and for the boundary sinh-Gordon
model \cite{key-6} are known. The bulk part of these theories can
be supersymmetrized in the framework described above, so we apply
our results to them.

The layout of the paper is as follows. In section 2 we review the
formalism for supersymmetric factorized scattering theory in the bulk
as described in \cite{key-3}, \cite{key-1} and \cite{key-15}. This
review is not a mere copy of results as we found the thorough understanding
of this formalism necessary for studying the boundary case. We hope
that we make certain points clearer, and we also supply certain details
(concerning representation theory for instance) that cannot be found
in \cite{key-3,key-1,key-15}. (The first part of section \ref{4.3}
is also relevant for the bulk formalism.) In section 3 we discuss
the boundary supersymmetry algebra and its action on the ground state
and multi-particle states, the boundary version of the Ansatz for
supersymmetric scattering theory, and the supersymmetric one-particle
ground state reflection factors for kinks and boson/fermion doublets.
Section 4 contains a description of the boundary supersymmetric bootstrap
structure and fusing rules under the condition that the bulk particles
transform in the kink and boson/fermion representations and the ground
state is a singlet with RSOS label $\hlf$. This section and especially
subsection \ref{4.3} contains the main results of the paper also
mentioned above. In section 5 we apply the general results of the
previous sections to specific models. We present our conclusions in
section 6. The Appendix contains the normalization factors for the
S-matrix blocks.

\section{Bulk supersymmetric factorized scattering and bootstrap\label{bulk}}

\begin{figure}
\begin{center}\includegraphics[%
  width=0.40\columnwidth]{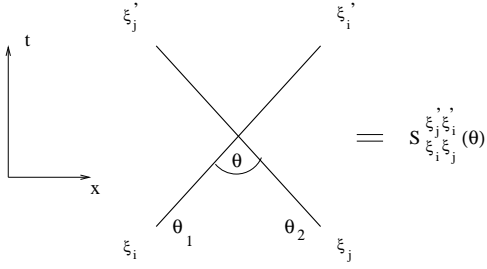}\end{center}

\caption{Two-particle S-matrix\label{fig smatrix}}
\end{figure}

The various particles contained in multi-particle \textsl{in, out}
or \textsl{intermediate} states will be written in their spatial order.
The order of the indices of the S-matrices will also be the same as
the spatial order of the corresponding particles (see figure \ref{fig smatrix}).
On the figures time flows upward. Upper indices of the S-matrix will
correspond to \textsl{out} states: $\ket c(\theta_{1})d(\theta_{2})_{in}=S_{cd}^{ab}(\theta_{1}-\theta_{2})\ket a(\theta_{2})b(\theta_{1})_{out}$.
If the S-matrix is written in matrix form (i.e. as a table of entries),
then the upper indices specify the rows and the lower indices specify
the columns. These conventions agree with \cite{key-3}. The {}``physical
strip'' for the rapidity argument $\theta$ of the S-matrix is $0<Im(\theta)<\pi$.%

Throughout the paper we mainly deal with the supersymmetric factors
of the S-matrix and with the supersymmetric part $\ket{A_{i}(\theta)}$
of the states. We shall usually refer to $\ket{A_{i}(\theta)}$ itself
as a state or supersymmetric state.

\subsection{Bulk supersymmetry charges\label{sec 2.1}}

The generators of the supersymmetry algebra are denoted by $N$, $Q$,
$\bar{Q}$, $\Gamma$. $Q$ and $\bar{Q}$ are called supercharges,
$\Gamma$ is the fermionic parity operator and $N$ is the boost generator.
The generators $Q,$ $\bar{Q},$ $\Gamma$ satisfy the relations \begin{equation}
\Gamma^{2}=1,\qquad\{\Gamma,Q\}=\{\Gamma,\bar{Q}\}=0.\label{susy1}\end{equation}
 The translation generators $H$ and $P$ are related to the supercharges:
\begin{equation}
Q^{2}=H+P,\qquad\bar{Q}^{2}=H-P.\label{susy2}\end{equation}
 It follows that the SUSY central charge $\hat{Z}=\hlf\{ Q,\bar{Q}\}$
commutes with $Q,$ $\bar{Q}$, $\Gamma$, $H$ and $P$. There are
also the relations\begin{equation}
[N,H+P]=H+P,\quad[N,H-P]=-(H-P),\quad[N,Q]=\hlf Q,\quad[N,\bar{Q}]=-\hlf\bar{Q}.\label{susy3}\end{equation}
 The coproduct $\Delta$ used to define the action of the supersymmetry
algebra on multi-particle states is given by\begin{equation}
\Delta(Q)=Q\otimes I+\Gamma\otimes Q,\quad\Delta(\bar{Q})=\bar{Q}\otimes I+\Gamma\otimes\bar{Q},\label{bulk copr1}\end{equation}
 \[
\Delta(\Gamma)=\Gamma\otimes\Gamma,\quad\Delta(Z)=\hat{Z}\otimes I+I\otimes\hat{Z},\]
 \[
\Delta(H)=H\otimes I+I\otimes H,\quad\Delta(P)=P\otimes I+I\otimes P,\quad\Delta(N)=N\otimes I+I\otimes N.\]
 On a one-particle supersymmetric multiplet $\ket{A_{i}}$\[
Q\ket{A_{i}(\theta)}=\sqrt{m_{i}}e^{\theta/2}q\ket{A_{i}(\theta)},\quad\bar{Q}\ket{A_{i}(\theta)}=\sqrt{m_{i}}e^{-\theta/2}\bar{q}\ket{A_{i}(\theta)},\]
 where $m_{i}$ is the mass of the multiplet and $q$ and $\bar{q}$
are $\theta$-independent matrices which act on the states of the
supermultiplet and satisfy\[
q^{2}=1,\qquad\bar{q}^{2}=1,\qquad\{ q,\bar{q}\}=2Z,\]
 where $Z=\frac{1}{m_{i}}\hat{Z}$ on the multiplet. The action of
$\Gamma$ is independent of $\theta$ and \[
\{\Gamma,q\}=\{\Gamma,\bar{q}\}=0.\]

\subsection{Particle and kink representations\label{sec 2.2}}

Two representations of the supersymmetry algebra are used to construct
supermultiplets: the boson-fermion and the kink representation. The
boson-fermion representation will often be referred to as the particle
representation. It contains a boson $\ket{\phi}$ and a fermion $\ket{\psi}$.
In the basis $\{\phi,\psi\}$\begin{equation}
q=\left(\begin{array}{cc}
0 & \epsilon\\
\epsilon^{*} & 0\end{array}\right),\quad\bar{q}=\left(\begin{array}{cc}
0 & \epsilon^{*}\\
\epsilon & 0\end{array}\right),\quad\Gamma=\left(\begin{array}{cc}
1 & 0\\
0 & -1\end{array}\right),\label{partrep}\end{equation}
 where $\epsilon=\exp(i\pi/4).$ The central charge is zero. Charge
conjugation $C$ acts as the identity: $\phi\leftrightarrow\phi,$
$\psi\leftrightarrow\psi.$ We denote this representation by $P$.
We will often refer to bosons and fermions simply as particles.

Another representation is obtained if we multiply $\Gamma$ in (\ref{partrep})
by $-1$. We call it pseudoparticle representation and denote it by
$\bar{P}$. The following equations describe the decomposition of
two-particle states: \begin{equation}
P\otimes P=P\oplus\bar{P},\quad\bar{P}\otimes\bar{P}=P\oplus\bar{P},\quad P\otimes\bar{P}=P\oplus\bar{P}.\label{cg1}\end{equation}
 In (\ref{cg1}) the third equation means for example that a two-particle
state containing a particle and a pseudoparticle transforms in the
sum of a particle and a pseudoparticle representation (of appropriate
mass).

The kink representation $K$ contains four kinks, interpolating between
three vacua labeled by the RSOS labels $0,\hlf,1.$ In the basis $\{ K_{0\hlf},K_{1\hlf},K_{\hlf0},K_{\hlf1}\}$\begin{equation}
q=\left(\begin{array}{cccc}
0 & i & 0 & 0\\
-i & 0 & 0 & 0\\
0 & 0 & 1 & 0\\
0 & 0 & 0 & -1\end{array}\right),\quad\bar{q}=\left(\begin{array}{cccc}
0 & i & 0 & 0\\
-i & 0 & 0 & 0\\
0 & 0 & -1 & 0\\
0 & 0 & 0 & 1\end{array}\right),\quad\Gamma=\left(\begin{array}{cccc}
0 & 1 & 0 & 0\\
1 & 0 & 0 & 0\\
0 & 0 & 0 & 1\\
0 & 0 & 1 & 0\end{array}\right).\label{kinkrep1}\end{equation}
 The central charge is\[
Z=\left(\begin{array}{cccc}
1 & 0 & 0 & 0\\
0 & 1 & 0 & 0\\
0 & 0 & -1 & 0\\
0 & 0 & 0 & -1\end{array}\right).\]
 There is also another representation $\bar{K}$ called pseudo-kink
representation, which is obtained by multiplying $q$ and $\bar{q}$
by $-1$ in (\ref{kinkrep1}), or by interchanging the RSOS labels
$0\leftrightarrow1$. Multi-kink states have to respect an adjacency
condition: in the physical \\
 $\ket{...K_{ab}(\theta_{1})K_{cd}(\theta_{2})...}$ state $b=c$
must hold. This condition gives kinks a character rather different
from that of particles. (The condition also implies that $(\Gamma\otimes I)\ket{K_{\hlf1}(\theta_{1})K_{1\hlf}(\theta_{2})}=0$,
for example.) Charge conjugation acts as follows: $K_{0\hlf}\leftrightarrow K_{\hlf0},$
$K_{1\hlf}\leftrightarrow K_{\hlf1}$. Multi-kink states containing
even number of kinks can be arranged in two sectors: the first sector
contains the states which have left and right RSOS index $\hlf$,
the second sector contains the states which have left and right RSOS
indices $0$ or $1$. These two sectors will be called $\hlf$ and
$01$ sector. For two-kink states we have \begin{equation}
K\otimes K=[P]_{\hlf}\oplus[P\oplus\bar{P}]_{01}.\label{cg3}\end{equation}
 The subscripts refer to the sectors in which the subspaces lie.

The $\otimes$ symbol in (\ref{cg3}) and further on denotes a tensor
product with the modification that the adjacency conditions (if there
are any) are also understood to be imposed. Thus $\otimes$ will sometimes
stand for a non-free tensor product, but this will not be denoted
explicitly.

The following combinations of two-kink states span the invariant subspaces
(see also \cite{key-15,key-11}):\begin{equation}
\ket{\phi_{1}}=\ket{K_{\hlf0}K_{0\hlf}}+\ket{K_{\hlf1}K_{1\hlf}},\qquad\ket{\psi_{1}}=\ket{K_{\hlf0}K_{0\hlf}}-\ket{K_{\hlf1}K_{1\hlf}};\label{frm1}\end{equation}
\begin{equation}
\ket{\phi_{2}}=\ket{K_{0\hlf}K_{\hlf0}}+\ket{K_{1\hlf}K_{\hlf1}},\qquad\ket{\psi_{2}}=\ket{K_{1\hlf}K_{\hlf0}}-\ket{K_{0\hlf}K_{\hlf1}};\label{frm2}\end{equation}
\begin{equation}
\ket{\bar{\phi}}=\ket{K_{0\hlf}K_{\hlf0}}-\ket{K_{1\hlf}K_{\hlf1}},\qquad\ket{\bar{\psi}}=\ket{K_{1\hlf}K_{\hlf0}}+\ket{K_{0\hlf}K_{\hlf1}}.\label{frm3}\end{equation}
 The rapidities do not play essential role in these formulae, so they
are suppressed. The two states in (\ref{frm3}) span the pseudoparticle
representation. The value of $\Gamma$ on $\ket{\bar{\phi}}$ is $-1$.
Although the decomposition (\ref{cg3}) is possible, the products
of elements of $[P]_{\hlf}$, $[P]_{01}$ and $[\bar{P}]_{01}$ satisfy
certain relations because of the kink adjacency conditions. For example,
$\ket{\bar{\phi}\bar{\phi}}=\ket{\phi_{2}\phi_{2}}$.

The eight particle-kink states $\ket{p(\theta_{1})k(\theta_{2})}$,
where $p$ stands for a boson or fermion and $k$ stands for a kink,
transform in the direct sum of a kink and a pseudo-kink representation
if and only if \begin{equation}
\theta_{1}-\theta_{2}=i(\pi-u)\quad{\textrm{and}}\quad m=2M\cos(u)=2M\sin(\pi/2-u),\label{kp fusing}\end{equation}
 where $m$ is the mass of the particle and $M$ is the mass of the
kink. This is precisely the condition that the total mass of the particle-kink
state is also $M$. If this condition is not satisfied, then the decomposition
of the representation in which the particle-kink states transform
does not contain the kink, pseudo-kink, particle or pseudoparticle
representations.

\subsection{Scattering amplitudes and supersymmetric bootstrap\label{sec 2.3}}

The general solution of the Yang-Baxter equations that describes the
(supersymmetric factor of the) scattering of particle supermultiplets
is {\small \begin{equation}
S_{P}^{[i,j]}(\theta)=G^{[i,j]}(\theta)\left[\frac{1}{2i}(q_{1}-q_{2})(\bar{q}_{1}-\bar{q}_{2})+\alpha F(\theta)[1-t(\theta,m_{i},m_{j})q_{1}q_{2}][1+t(\theta,m_{j},m_{i})\bar{q}_{1}\bar{q}_{2}]\right],\label{SP}\end{equation}
} where \[
t(\theta,m_{i},m_{j})=\tanh\left(\frac{\theta+\log(m_{i}/m_{j})}{4}\right),\qquad F(\theta)=\frac{m_{i}+m_{j}+2\sqrt{m_{i}m_{j}}\cosh(\theta/2)}{2i\sinh(\theta)},\]
 $q_{1}=q\otimes I,$ $q_{2}=\Gamma\otimes q,$ $\bar{q}_{1}=\bar{q}\otimes I,$
$\bar{q}_{2}=\Gamma\otimes\bar{q}$. $m_{i}$ and $m_{j}$ are the
masses of the multiplets, and $\alpha$ is a real constant, which
measures the strength of Bose-Fermi mixing (the dependence of $\alpha$
on $i,j$ is not indicated). $S_{P}^{[i,j]}(\theta)/G^{[i,j]}(\theta)$
can depend on the conserved quantities $i,j$ through $m_{i}$, $m_{j}$
and $\alpha$ only. It also follows from the Yang-Baxter equation
that the particles in a theory can be divided into disjoint sets with
the property that any two particles in a set have the same nonzero
$\alpha$, and $\alpha=0$ for two particles from different sets.
$\alpha=0$ corresponds to trivial scattering. To each particle in
a theory we associate a value of $\alpha$, which is the value that
occurs in the scattering of the particle with itself. For simplicity
we consider only theories which have only one such set and thus $\alpha$
is the same for any two-particle scattering. The scalar function $G^{[i,j]}(\theta)$
is determined by unitarity and crossing symmetry up to CDD factors.
It is important here that $i,j$ are invariant under charge conjugation.
It is also required that $S_{P}^{[i,j]}(\theta)$ should be minimal,
what fixes $G^{[i,j]}(\theta)$ completely. The minimal solution can
be found in the Appendix (cf. \cite{key-1,key-3}). $G^{[i,j]}(\theta)$
contains the parameters $u_{i}$, $u_{j}$ for which \begin{equation}
0<Re(u_{i}),Re(u_{j})\leq\pi/2,\qquad m_{i}=2M\sin(u_{i}),\quad m_{j}=2M\sin(u_{j}),\label{u-k}\end{equation}
 where $M=|1/(2\alpha)|$. Consequently, for each particle there is
a corresponding angle $u$. In this paper we consider only real values
of $u_{i}$ and $u_{j}$.

The S-matrix factor that describes the scattering of two kinks of
equal mass is\[
S_{K}^{[i,j]}(\theta)=K^{[i,j]}(\theta)[\cosh(\gamma\theta)-\sinh(\gamma\theta)q_{1}\bar{q}_{1}][\cosh(\theta/4)-\sinh(\theta/4)q_{1}q_{2}],\]
 where $\gamma=\log2/2\pi i.$ The scalar function $K^{[i,j]}(\theta)$
is determined by unitarity and crossing symmetry and the condition
that $S_{K}^{[i,j]}(\theta)$ should be minimal. See the Appendix
or \cite{key-1} for $K^{[i,j]}(\theta)$. $S_{K}^{[i,j]}(\theta)$
is bijective in the physical strip. The scattering of kinks of different
mass is impossible, so all the kinks in a theory have to have the
same mass. The kink-particle S-matrix $S_{PK}(\theta)$ will be considered
later. The upper indices $^{[i,j]}$ will often be suppressed.

The important common feature of these minimal S-matrices, including
$S_{PK},$ is that they have no poles and overall zeroes in the physical
strip (although they can be degenerate at particular values of $\theta$).
It is required that the scattering of all particles and kinks of the
supersymmetrized theory should be described by minimal supersymmetric
S-matrix factors, i.e. $S_{P},$ $S_{K}$ and $S_{PK}$ must be the
supersymmetric two-particle S-matrix building blocks. It turns out
that this requirement is consistent with the bootstrap equation, so
the fusing data of the non-supersymmetric theory are not modified
essentially by adding the supersymmetric S-matrix factors.

\begin{figure}
\begin{center}\includegraphics[%
  width=0.20\textwidth]{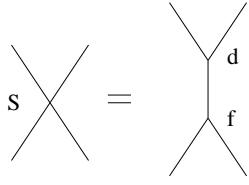}\end{center}

\caption{\label{fig bulksplit} The fusing equation (splitting of S)}
\end{figure}

The three-particle couplings (or fusing and decay tensors) are also
assumed to factorize into a non-supersymmetric and supersymmetric
part: $\hat{f}_{\xi_{i}A_{i}\xi_{j}A_{j}}^{\xi_{k}A_{k}}=\tilde{f}_{\xi_{i}\xi_{j}}^{\xi_{k}}f_{A_{i}A_{j}}^{A_{k}}$,
$\hat{d}_{\xi_{k}A_{k}}^{\xi_{i}A_{i}\xi_{j}A_{j}}=\tilde{d}_{\xi_{k}}^{\xi_{i}\xi_{j}}d_{A_{k}}^{A_{i}A_{j}}$,
and for the supersymmetric factor of the S-matrix \begin{equation}
S_{A_{i}A_{j}}^{A_{j}'A_{i}'}(iu_{ij}^{k})=\sum_{A_{k}}f_{A_{i}A_{j}}^{A_{k}}d_{A_{k}}^{A_{j}'A_{i}'}\label{fusing equ}\end{equation}
 holds, where $u_{ij}^{k}$ denotes the $\xi_{i}+\xi_{j}\rightarrow\xi_{k}$
fusing angle, and \begin{equation}
\ket{A_{i}(\theta+i(\pi-u_{ik}^{j}))A_{j}(\theta-i(\pi-u_{jk}^{i}))}=\sum_{A_{k}}f_{A_{i}A_{j}}^{A_{k}}\ket{A_{k}(\theta)},\label{fusing1}\end{equation}
\begin{equation}
\ket{A_{k}(\theta)}=\sum_{A_{i},A_{j}}d_{A_{k}}^{A_{i}A_{j}}\ket{A_{i}(\theta-i(\pi-u_{ik}^{j}))A_{j}(\theta+i(\pi-u_{jk}^{i}))},\label{fusing2}\end{equation}
 where $u_{ik}^{j},$ $u_{jk}^{i}$ are the other angles at the three-particle
vertices. We call these equations fusing equations. The $f$ fusing
tensors and the $d$ decay tensors should commute with the action
of supersymmetry. See fig. \ref{fig bulksplit} for graphical representation.%

Once the supersymmetry representations in which the various particles
transform are fixed, these equations can severely constrain the possible
fusing angles in a sense that we describe next, beginning with $particle+particle\rightarrow particle$
fusion.

In the light of (\ref{cg1}), if it is decided that some particles
and their bound states transform in the particle representation of
the SUSY algebra, then the fusing equation (see figure \ref{fig bulksplit})
can be satisfied if and only if $S_{P}(iu_{ij}^{k})$ is a projection
on the appropriate subspace carrying the particle representation.
This is a nontrivial condition on $S_{P}(iu_{ij}^{k})$, because $S_{P}(\theta)$
is bijective (of rank four) in general. The other possibility to assure
that only particles are produced in the fusion is to factor out the
unwanted states by hand. Note that because of (\ref{fusing1}) and
(\ref{fusing2}) this implies that certain two-particle \textsl{in}
and \textsl{out} states with complex rapidities also have to be factored
out. In the present paper we consider the first, more natural possibility.

$S_{P}^{[i,j]}(\theta)$ is of rank two if $\theta=iu_{ij}^{k}$,
where $u_{ij}^{k}\in\{ u_{i}+u_{j},\pi-u_{i}+u_{j},u_{i}+\pi-u_{j}\}$.
Only two of these values can be in the physical strip, and $S_{P}^{[i,j]}(\theta)$
is nondegenerate at other values of $\theta$ in the physical strip.
The image space of $S_{P}^{[i,j]}(iu_{ij}^{k})$ carries the particle
representation if and only if $\alpha<0$, i.e. if $\alpha=-1/(2M)$.
We remark that if $\alpha=1/(2M)$, then the image space carries the
pseudoparticle representation. The value of $u_{k}$ is the following:

\begin{equation}
u_{k}=u_{i}+u_{j}\quad\textrm{if}\quad u_{ij}^{k}=u_{i}+u_{j}<\pi/2,\label{angle 1}\end{equation}
\begin{equation}
u_{k}=\pi-(u_{i}+u_{j})\quad\textrm{if}\quad u_{ij}^{k}=u_{i}+u_{j}\geq\pi/2,\label{angle 2}\end{equation}
\begin{equation}
u_{k}=u_{i}-u_{j}\quad\textrm{if}\quad u_{ij}^{k}=\pi-u_{i}+u_{j},\label{angle 3}\end{equation}
\begin{equation}
u_{k}=u_{j}-u_{i}\quad\textrm{if}\quad u_{ij}^{k}=u_{i}+\pi-u_{j}.\label{angle 4}\end{equation}
 The other angles at the three-particle vertex are \[
u_{jk}^{i}=\pi-u_{i},\quad u_{ki}^{j}=\pi-u_{j}\quad\textrm{if}\quad u_{ij}^{k}=u_{i}+u_{j},\]
\begin{equation}
u_{jk}^{i}=u_{i},\quad u_{ki}^{j}=\pi-u_{j}\quad\textrm{if}\quad u_{ij}^{k}=\pi-u_{i}-u_{j},\end{equation}
 \[
u_{jk}^{i}=\pi-u_{i},\quad u_{ki}^{j}=u_{j}\quad\textrm{if}\quad u_{ij}^{k}=u_{i}+\pi-u_{j}.\]
 The fusions corresponding to (\ref{angle 3}) or (\ref{angle 4})
are crossed versions of (\ref{angle 1}), (\ref{angle 2}). Explicit
expressions for the three-particle couplings are written down in \cite{key-3,key-1}.%
\begin{figure}
\begin{center}\includegraphics[%
  width=0.25\textwidth]{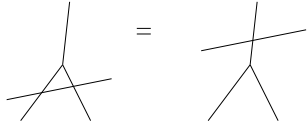}\end{center}

\caption{\label{fig bstrbulk}Bootstrap equation}
\end{figure}

As discussed above, we require the minimality of the supersymmetric
two-particle S-matrices. This is yet another nontrivial condition
for the possible fusions, because the bootstrap equation determines
the scattering of the produced particle on other particles. It is
clear that the (two-particle) supersymmetric S-matrix given by the
bootstrap equation will be $S_{P}$ with $\alpha=-1/(2M)$ modulo
CDD factors, because a two-particle S-matrix given by the bootstrap
equation automatically satisfies the axioms of factorized scattering,
i.e. the Yang-Baxter, unitarity and crossing equations, and real analyticity,
and we know that the solution to these constraints is $S_{P}$ modulo
CDD factors. By checking the pole structure it can be seen that the
(supersymmetric) particle S-matrix given by bootstrap is in fact also
$S_{P}$, so the minimality condition is consistent with the bootstrap
and does not give more constraints on the fusion of (supersymmetric)
particles.

We turn to the case of the fusion of two (supersymmetric) kinks of
equal mass now. The kink scattering amplitude is always of rank six
(bijective) in the physical strip. Consequently, in the light of (\ref{cg3}),
if one insists that no pseudoparticles should be formed in kink fusion,
then one has to factor out the unwanted states from the Hilbert-space
by hand. Because of (\ref{fusing1}) and (\ref{fusing2}) this implies
that certain two-kink \textsl{in} and \textsl{out} states with complex
rapidities also have to be factored out. Even if states of the form
(\ref{frm3}) with kink rapidities $\theta+iu$ and $\theta-iu$,
$2u$ being the fusing angle, are factored out, kink fusion produces
two types of particles corresponding to (\ref{frm1}) and (\ref{frm2}),
i.e. to the $\hlf$ and $01$ sectors. The two types of particles
will be referred to as type $\hlf$ and type $01$ particles.

There are adjacency conditions for particles produced in kink fusion,
which follow from the adjacency conditions for kinks: type $\hlf$
and type $01$ particles cannot be adjacent in a multi-particle state,
so they cannot scatter on each other. There are also appropriate adjacency
conditions for kinks and particles. Bootstrap gives the result \cite{key-15,key-1}
that the two types of particles have the same S-matrix $S_{P}$. Consequently,
the two types of particles can be identified (which is the same as
factoring out certain combinations). If this identification is made,
then only the following adjacency condition applies: if in a state
$\ket{...K_{ab}p...p...pK_{cd}...}$ (where $p$ stands for a particle
and the rapidities are suppressed) there are only particles between
$K_{ab}$ and $K_{cd}$, then either $b=c$, or $|b-c|=1$.

In this paper we restrict ourselves to theories in which pseudoparticles
do not occur (see also section \ref{4.3}). Consequently, we have
to factor out the pseudoparticles that would arise. We also identify
the two types of particles%
\footnote{The same factorizations are done in the literature.%
}.

There is no condition on the rank of the kink-kink scattering amplitude
$S_{A_{i}A_{j}}^{A_{j}'A_{i}'}(iu_{ij}^{k})$, and so the fusing equations
can be satisfied for arbitrary fusing angles. The mass of the resulting
state will be \[
m_{3}=2M\cos(u_{ij}^{k}/2)=2M\sin(\pi/2-u_{ij}^{k}/2),\]
 where $M$ is the mass of the kinks. The fusing tensor as linear
map is bijective for the $kink+kink\rightarrow particle$ fusion.

The S-matrix $S_{PK}(\theta)$ for the scattering of a particle with
$\alpha<0$ and a kink of mass $M=-1/(2\alpha)$ is obtained from
$S_{K}$ by bootstrap \cite{key-1,key-15} applied to the $kink+kink\rightarrow particle$
vertex. It turns out that $S_{PK}(\theta)$ is also minimal and has
neither poles nor zeroes in the physical strip.

It is expected that a kink is produced in the kink-particle fusion.
The transformation properties of the kink-particle states discussed
earlier show that in this case it is necessary that (\ref{kp fusing})
is satisfied. We checked that $S_{PK}$ is bijective (of rank eight)
everywhere in the physical strip, except (\ref{kp fusing}) is satisfied.
In the latter case it is a projection onto the four dimensional kink
subspace. The $kink+particle\rightarrow kink$ fusion is thus possible
indeed, and there are no restrictions other than (\ref{kp fusing}).
The $kink+particle\rightarrow kink$ fusion is a crossed version of
$kink+kink\rightarrow particle$ fusion. The produced kink is of the
same mass as the incoming one, so the fusing angle is in the domain
$[\pi/2,\pi].$ The fusing tensor is a projection.

Finally, there are bootstrap equations for $S_{P}$, $S_{K},$ $S_{PK}$
and the fusing processes described above, which are expected to be
valid. They were checked, and the result is \cite{key-1,key-3} that
they are indeed satisfied: the scattering amplitudes that can be obtained
from $S_{K}$, $S_{P}$ and $S_{PK}$ by bootstrap are $S_{K}$, $S_{P}$,
and $S_{PK}$ again. In particular, the condition of minimality is
consistent with the bootstrap. The fusion of two particles with $\alpha<0$
produces a particle with the same value of $\alpha$. The fusion of
two kinks of mass $M$ produces a particle with $\alpha=-1/(2M)$,
and if the fusing angle is $\rho$, then the angle $u$ corresponding
to the produced particle is $u=\pi/2-\rho/2$. The fusion of a kink
of mass $M$ and a particle with $\alpha=-1/(2M)$ produces a kink
of mass $M$. %
\begin{figure}
\begin{center}\includegraphics[%
  width=0.10\textwidth]{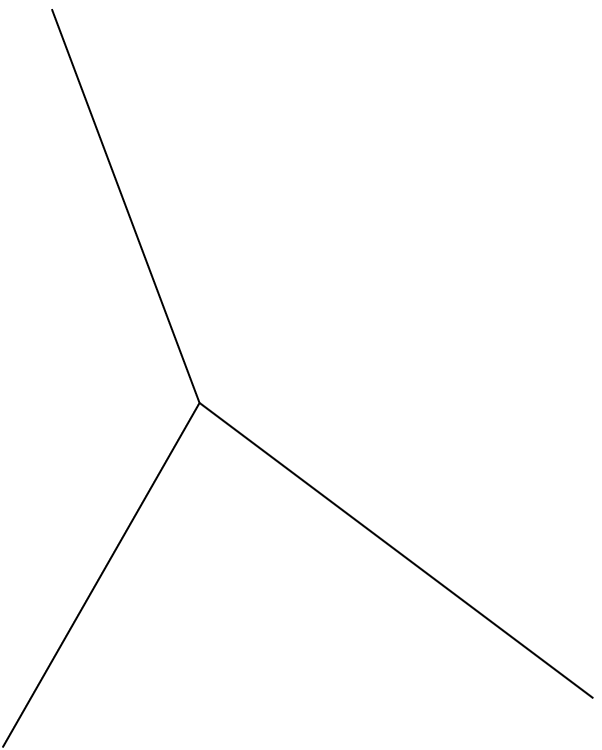}\qquad\qquad \includegraphics[%
  width=0.10\textwidth]{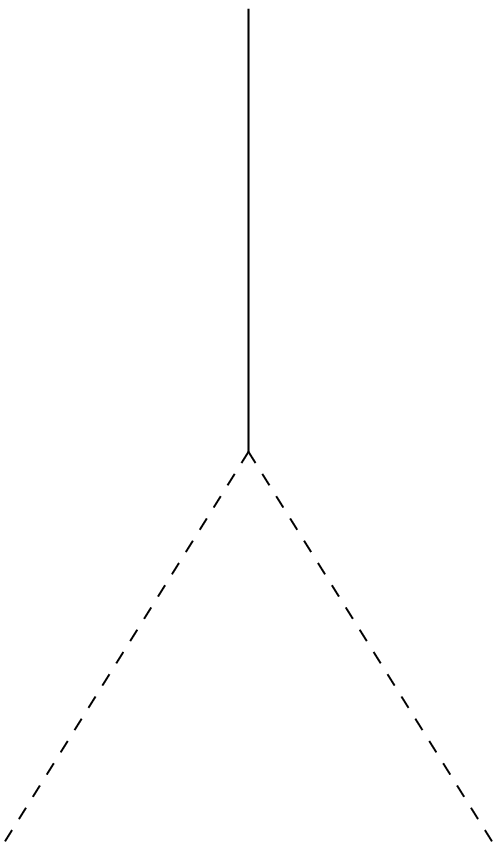}\end{center}

\caption{\label{fig vertexek}The two bulk vertices}
\end{figure}

The three-particle and the kink-kink-particle vertices are shown in
figure \ref{fig vertexek} (the crossed versions of them are also
possible). Figure \ref{fig vertexek} is the only one in which we
use different line styles for different types of particles.

\section{Boundary supersymmetric factorizable scattering}

Let us assume that a bulk factorizable scattering theory and a supersymmetric
version of this theory are given, and the supersymmetric version is
of the form described in section 2.

In particular, the bulk supersymmetric theory is characterized by
a single $\alpha<0$ (if there are particles in the theory) for the
scattering of any two particles, and a kink mass $M>0$ (if there
are kinks in the theory). $\alpha=-1/(2M)$ is satisfied if there
are both kinks and particles in the theory. Each particle has mass
$m\leq-1/\alpha=2M$, and to each particle an angle $0<u\leq\pi/2$
is associated as described in the previous section (see (\ref{u-k})).
The fusing rules of the bulk theory satisfy the constraints described
in section 2. These constraints are $u_{ij}^{k}\in\{ u_{i}+u_{j},\pi-u_{i}+u_{j},u_{i}+\pi-u_{j}\}$
for the fusing angle of a $particle_{i}+particle_{j}\rightarrow particle_{k}$
process, $u_{k}=\pi/2-u_{ij}^{k}/2$ for a $kink_{i}+kink_{j}\rightarrow particle_{k}$
process, and $u_{ij}^{k}=\pi/2+u_{i}$ for a $particle_{i}+kink_{j}\rightarrow kink_{k}$
process.

It is also assumed that a factorizable boundary scattering theory
associated to the non-supersymmetric bulk theory is also given, and
the boundary is on the right hand side.

\subsection{Boundary supersymmetry algebra}

The presence of boundary partially violates supersymmetry. In general
only one supercharge $\tilde{Q}$ and a central charge $\tilde{Z}$
is expected to be conserved if a supersymmetric boundary is present.
We call the boundary version of supersymmetry boundary supersymmetry.
The boundary supersymmetry algebra is denoted by $\mathcal{A}_{by}$,
and the full bulk supersymmetry algebra by $\mathcal{A}_{bulk}$.

In any factorizable boundary scattering theory there is a Hilbert-space
$\mathcal{H}_{B}$ of the boundary bound states (also containing the
ground state(s)). General multi-particle \textsl{in} states containing
$n\geq1$ bulk particles traveling towards the boundary form a space
$\mathcal{H}_{n,by}=\mathcal{H}_{n,b}\otimes\mathcal{H}_{B}$, where
$\mathcal{H}_{n,b}$ is a subspace of the Hilbert-space $\mathcal{H}_{n}$
of bulk $n$-particle \textsl{in} states. $\mathcal{H}_{n,b}$ is
the subspace containing \textsl{in} particles traveling towards the
boundary. The Hilbert-space of the boundary theory is thus $\mathcal{H}_{by}=\mathcal{H}_{B}\oplus(\oplus_{n=1}^{\infty}\mathcal{H}_{n,by})$.
In the supersymmetric situation a representation $\pi_{1}$ of $\mathcal{A}_{bulk}$
is defined on the Hilbert-space of the bulk theory, so in order to
define the action of $\mathcal{A}_{by}$ on $\mathcal{H}_{by}$ we
need the action $\pi_{2}$ of $\mathcal{A}_{by}$ on $\mathcal{H}_{B}$,
and a coproduct $\Delta_{B}:\mathcal{A}_{by}\rightarrow\mathcal{A}_{bulk}\otimes\mathcal{A}_{by}$.
$\mathcal{A}_{by}$ itself need not admit a coproduct. We also require
the coassociativity property $(Id\otimes\Delta_{B})\Delta_{B}=(\Delta\otimes Id)\Delta_{B}$,
so $\mathcal{A}_{by}$ is a coideal of $\mathcal{A}_{bulk}$. A similar
construction applies to the action of $\mathcal{A}_{by}$ on \textsl{out}
states and \textsl{intermediate} multi-particle states. This framework
is the same as that used in \cite{key-13,key-14} for quantum group
symmetry, and it is consistent with the definition of the action of
supersymmetry in \cite{key-2,key-9}. In fact, we need to specify
the action of $\mathcal{A}_{by}$ on the ground state(s) only, because
its action on the other elements of $\mathcal{H}_{B}$ is determined
by the boundary bootstrap. We do not require in general that the action
of the supersymmetry generators $Q$, $\bar{Q}$, and $\Gamma$ should
be defined on the boundary ground state(s) or on other elements of
$\mathcal{H}_{B}$. We remark that in the construction described for
instance in \cite{key-2} the coproduct $\Delta$ is used, so it is
necessary to specify the action of $Q$, $\bar{Q}$, and $\Gamma$
on the ground state. $\mathcal{A}_{by}$ is a subalgebra of $\mathcal{A}_{bulk}$
in that construction, and the reflection amplitudes are invariant
only with respect to $\mathcal{A}_{by}$.

In the following paragraphs we specify $\mathcal{A}_{by}$ and $\Delta_{B}$.
Our formulae are not derived from Lagrangian field theory in a rigorous
manner, neither do they result from some rigorous classification of
all coideals of $\mathcal{A}_{bulk}$, but they are consistent with
considerations and results in \cite{key-2,key-13,key-14,key-8,key-9},
and we give further arguments supporting them. We think that the framework
we describe is fairly general, and the assumptions we make are not
really restrictive.

$\mathcal{A}_{by}$ is assumed to have the generators $\tilde{Q}$,
$\tilde{H}$, $\tilde{Z}$ and $\tilde{I}$. $\tilde{I}$ is the identity
element, $\tilde{H}$ is the time translation generator. The conservation
of fermionic parity is not required, so the corresponding element
$\tilde{\Gamma}$ is not included in $\mathcal{A}_{by}$. The boundary
supersymmetric central charge $\tilde{Z}$ commutes with the other
generators. The supercharge $\tilde{Q}$ is expected to be the boundary
counterpart of $Q$ and $\bar{Q}$, so we expect that $\tilde{Q}^{2}$
is of the form\begin{equation}
\tilde{Q}^{2}=2\tilde{H}+p(\tilde{I},\tilde{Z}),\label{rel}\end{equation}
 where $p$ is a polynomial. This relation implies that $\mathcal{A}_{by}$
is a commutative algebra. It is natural to assume that in analogy
with the bulk SUSY algebra \begin{equation}
\Delta_{B}(\tilde{H})=H\otimes\tilde{I}+I\otimes\tilde{H}\quad{\textrm{and}}\quad\Delta_{B}(\tilde{Z})=\hat{Z}\otimes\tilde{I}+I\otimes\tilde{Z}.\label{copr1}\end{equation}
 Because of the expected role of $\tilde{Q}$ and the relation (\ref{rel})
for $\tilde{Q}^{2}$ we take the Ansatz $\Delta_{B}(\tilde{Q})=Q\otimes X_{1}+\bar{Q}\otimes X_{2}+\Gamma\otimes X_{3}+I\otimes X_{4}+\hat{Z}\otimes X_{5}$.
Solving the coassociativity condition and (\ref{rel}) for the $X_{i}$-s
we found that \begin{equation}
\Delta_{B}(\tilde{Q})=(Q\pm\bar{Q})\otimes\tilde{I}+\Gamma\otimes\tilde{Q},\label{copr2}\end{equation}
\begin{equation}
\tilde{Q}^{2}=2\tilde{H}\pm2\tilde{Z}+2u\tilde{I},\label{qbnegyzet}\end{equation}
 where $u$ is an arbitrary number. Two cases can be distinguished
corresponding to the sign in (\ref{copr2}) and (\ref{qbnegyzet}),
they will be referred to as $(+)$ and $(-)$ cases, respectively.
$u$ can be tuned by the redefinition of $\tilde{H}$ : $\tilde{H}\rightarrow\tilde{H}+u'\tilde{I}$
results in $u\rightarrow u-u'$. $u$ can also be tuned by a similar
redefinition of $\tilde{Z}$. We set $u$ equal to zero by the redefinition
$\tilde{H}+u\tilde{I}\rightarrow\tilde{H}$: \begin{equation}
\tilde{Q}^{2}=2\tilde{H}\pm2\tilde{Z}.\label{qbnegyzet2}\end{equation}

Another way to obtain the formula (\ref{copr2}) is the following:
let us adopt first that $\Delta_{B}(\tilde{Q})=(Q\pm\bar{Q})\otimes\tilde{I}+a\, boundary\, contribution$,
i.e. $\tilde{Q}$ is the boundary counterpart of $Q\pm\bar{Q}$. This
is supported by considerations in the Lagrangian framework \cite{key-2,key-9,key-8}.
It is reasonable \cite{key-2} now to adopt (\ref{qbnegyzet2}) as
the boundary version of the bulk formula $(Q\pm\bar{Q})^{2}=2H\pm2\hat{Z}$.
Now $\mathcal{A}_{by}$ is isomorphic to a subalgebra $\mathcal{A}_{by}'$
of $\mathcal{A}_{bulk}$, the isomorphism $i$ is given by $\tilde{Q}\mapsto Q\pm\bar{Q}$,
$\tilde{H}\mapsto H$, $\tilde{Z}\mapsto\hat{Z}$. If $\Delta(\mathcal{A}_{by}')\subset\mathcal{A}_{bulk}\otimes\mathcal{A}_{by}'$,
which can easily be verified to be true, then we can define $\Delta_{B}$
using $\Delta$ as follows: $\Delta_{B}(X)=(Id\otimes i)^{-1}\Delta(i(X))$,
$X\in\mathcal{A}_{by}$. This definition guarantees coassociativity,
and it is easy to check that it leads to (\ref{copr2}), (\ref{copr1}).
If we took (\ref{qbnegyzet}) with nonzero $u$, then $i(\tilde{Q})=Q\pm\bar{Q}+\sqrt{u}\Gamma$
would be an appropriate choice. The formulae (\ref{copr2}), (\ref{copr1})
are independent of $u$.

The boundary supersymmetry algebra as described above is isomorphic
to a coideal subalgebra of $\mathcal{A}_{bulk}$. This situation and
the construction above for the boundary supersymmetry algebra is very
similar to that described in \cite{key-13,key-14} for quantum group
symmetry.

$\Delta_{B}(\tilde{Q})$, $\Delta_{B}(\tilde{H})$ and also $\Delta_{B}(\tilde{Z})$
is a sum of two terms. The first terms can be regarded as the bulk
parts of these quantities, and the second terms can be regarded as
boundary contributions. This structure is in accord with the expectation
that the quantities $\tilde{Q}$, $\tilde{H}$, $\tilde{Z}$ are (semi)local
in the underlying Lagrangian theory. Locality also implies that the
bulk and the boundary contributions commute, which is also true for
our formulae (see also \cite{key-2}).

If in a certain model the ground state is a singlet, then we have\begin{equation}
\tilde{Q}\ket{B}=\gamma\ket{B},\quad\tilde{Z}\ket{B}=z\ket{B},\label{vakuumabrazolas}\end{equation}
 where $\ket{B}\in\mathcal{H}_{B}$ is the ground state, $\gamma$
and $z$ are numbers ($z\in\RR$). For the Hamiltonian operator \[
\tilde{H}\ket{B}=h\ket{B}=((\gamma^{2}/2)\mp z)\ket{B}\]
 holds if $u=0$. We expect that only $\gamma$ is a true parameter
of the model. $\gamma$ is expected to be expressible in terms of
the parameters of the bulk and boundary parts of the underlying classical
Lagrangian density. On the subspace $(\oplus_{n=1}^{\infty}\mathcal{H}_{n,by})\otimes\ket{B}$
the generators are $\tilde{Q}=Q\otimes I\pm\bar{Q}\otimes I+\gamma\Gamma\otimes I$
(cf. \cite{key-2}), $\tilde{Z}=\hat{Z}\otimes I+zI\otimes I,$ $\tilde{H}=H\otimes I+hI\otimes I$.
(The representations $\pi_{1}$, $\pi_{2}$ are not explicitely designated.)

It is not surprising that $\mathcal{A}_{by}$ is isomorphic to a subalgebra
of $\mathcal{A}_{bulk}$, because $\mathcal{A}_{by}$ acts on the
singlet $\ket{B}$ by multiplications, so $\pi_{3}(\mathcal{A}_{by})$,
where $\pi_{3}$ is the representation of $\mathcal{A}_{by}$ on $(\oplus_{n=1}^{\infty}\mathcal{H}_{n,by})\otimes\ket{B}$,
is a subalgebra of $\pi_{2}(\mathcal{A}_{bulk})\otimes I$.

A third way to obtain (\ref{copr2}) and (\ref{qbnegyzet}) is the
following: consider the representation of $\mathcal{A}_{by}$ on $(\oplus_{n=1}^{\infty}\mathcal{H}_{n,by})\otimes\ket{B}$.
Using (\ref{copr1}) and (\ref{vakuumabrazolas}) we have $\tilde{Z}=\hat{Z}\otimes I+zI\otimes I,$
$\tilde{H}=H\otimes I+hI\otimes I$ on $(\oplus_{n=1}^{\infty}\mathcal{H}_{n,by})\otimes\ket{B}$.
Considering (\ref{rel}) and (\ref{vakuumabrazolas}) we take the
Ansatz $\tilde{Q}=aQ\otimes I+b\bar{Q}\otimes I+c\Gamma\otimes I+dI\otimes I+e\hat{Z}\otimes I$
on $(\oplus_{n=1}^{\infty}\mathcal{H}_{n,by})\otimes\ket{B}$, where
$a,b,c,d,e$ are numbers. One can now solve (\ref{rel}) for $a,b,c,d,e$.
The result is $\tilde{Q}=Q\otimes I\pm\bar{Q}\otimes I+\gamma\Gamma\otimes I$
and the formula (\ref{qbnegyzet}), with $h=\gamma^{2}/2\mp z-u$.
The precise structure of $\mathcal{A}_{by}$ is thus obtained and
the algebra is also realized as a subalgebra of $\mathcal{A}_{bulk}$.
$\Delta_{B}(\tilde{Q})$ can be derived from $\Delta$ as earlier,
or it can be found by using coassociativity and the property that
$\Delta_{B}$ is an algebra homomorphism.

To describe situations when the fermionic parity is also conserved,
$\mathcal{A}_{by}$ can be supplemented by the boundary fermionic
parity generator $\tilde{\Gamma}$. It satisfies the following relations:\begin{equation}
\tilde{\Gamma}^{2}=\tilde{I},\quad[\tilde{\Gamma},\tilde{Z}]=0,\quad[\tilde{\Gamma},\tilde{H}]=0,\label{bgamma}\end{equation}
 and also \begin{equation}
\{\tilde{\Gamma},\tilde{Q}\}=2g\tilde{I},\label{bgamma2}\end{equation}
 where $g$ is a parameter. The coproduct of $\tilde{\Gamma}$ is
\begin{equation}
\Delta_{B}(\tilde{\Gamma})=\Gamma\otimes\tilde{\Gamma}.\label{copr3}\end{equation}
 Formula (\ref{bgamma2}) can be obtained from the requirement that
$\Delta_{B}$ is a homomorphism.

In case of a model with singlet boundary ground state \[
\tilde{\Gamma}\ket{B}=\epsilon\ket{B},\qquad\epsilon=\pm1,\]
 and $g=\epsilon\gamma$. The sign $\epsilon$ is not expected to
be a true parameter of the model (see also \cite{key-2}).

\subsection{Supersymmetrized reflection amplitudes}

We assume that ---similarly to the bulk case--- the boundary states
$\ket{\eta_{i}}$ of the non-supersymmetric theory become multiplets
$\ket{B_{i}}\otimes\ket{\eta_{i}}$ in the supersymmetrized boundary
theory, and that the one-particle reflection amplitude $\hat{R}$
on these states take the form\begin{equation}
\hat{R}_{\xi_{i}A_{i};\eta_{j}B_{j}}^{\xi_{i}'A_{i}';\eta_{j}'B_{j}'}(\theta)=R_{A_{i}B_{j}}^{A_{i}'B_{j}'}(\theta)\tilde{R}_{\xi_{i}\eta_{j}}^{\xi_{i}'\eta_{j}'}(\theta),\label{ansatz2}\end{equation}
 where $\tilde{R}$ is the one-particle reflection amplitude in the
non-supersymmetric model, and $R$ is called the supersymmetric factor.
$B_{i}$ specify the boundary supersymmetric quantum numbers, boundary
supersymmetry acts trivially on the non-supersymmetric indices. The
index $i$ specifies quantum numbers which are conserved in the non-supersymmetric
boundary theory. These quantum numbers should be defined and conserved
in the bulk theory as well, and in the bulk theory they should be
Poincare and charge conjugation invariant. This implies that the bulk
SUSY representations in which the particles transform should depend
only on quantum numbers which are also conserved in reflections. This
is a restriction on the bulk supersymmetric theory in principle. It
is also required that the reflection amplitude $\hat{R}$ and so also
$R$ commute with the boundary supersymmetry.

The factorized form of $\hat{R}$ (and $\hat{S}$) implies that the
boundary Yang-Baxter equations, the boundary unitarity equation and
the boundary cross-unitarity equation also factorize, so $R$ has
to satisfy them separately in order for $\hat{R}$ to satisfy them.
The boundary fusing equations and fusing tensors are also assumed
to factorize, and this implies that the boundary bootstrap equations
also factorize. The non-supersymmetric part of the bootstrap equations
are satisfied, so we require that the bootstrap equations should be
satisfied by the supersymmetric factors separately.

As in the bulk case, the non-supersymmetric boundary (and bulk) fusing
data enter the fusing and bootstrap equations for the supersymmetric
factors, and this places nontrivial conditions on the described supersymmetrization
scheme.

It is required that the supersymmetric factors of the reflection amplitudes
should be minimal, without poles and overall zeroes on the imaginary
axis in the {}``physical strip'', which is $0<Im(\theta)<\pi/2$
for the rapidity argument $\theta$ of the reflection amplitudes.
This guarantees that no new bound states are introduced by the supersymmetrization.

\begin{figure}
\begin{center}\includegraphics[%
  width=0.20\textwidth]{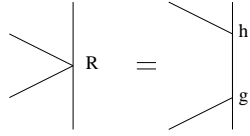}\end{center}

\caption{\label{fig brysplit}Boundary fusing equation (splitting of $R$)}
\end{figure}

The supersymmetric boundary fusing equations are analogous with the
bulk ones:

\[
R_{A_{i}B_{j}}^{A_{i}'B_{j}'}(i\nu_{ij}^{k})=\sum_{B_{k}}g_{A_{i}B_{j}}^{B_{k}}h_{B_{k}}^{A_{i}'B_{j}'},\qquad\ket{A_{i}B_{j}}=\sum_{B_{k}}g_{A_{i}B_{j}}^{B_{k}}\ket{B_{k}},\quad\ket{B_{k}}=\sum_{A_{i},B_{j}}h_{B_{k}}^{A_{i}B_{j}}\ket{A_{i}B_{j}},\]
 $\nu_{ij}^{k}$ is the boundary fusing angle for the $\xi_{i}$+$\eta_{j}\rightarrow\eta_{k}$
fusion, $g$ and $h$ are the boundary fusing and decay tensors respectively.
For graphical representation see fig. \ref{fig brysplit}.

\subsection{Ground state representation}

There are several different representations of the boundary supersymmetry
algebra which can serve as representations in which the ground states
of various theories transform. Moreover, it is not necessarily a natural
assumption that the supersymmetric part of the ground state of a certain
theory is a singlet. Nevertheless, in this paper we consider theories
which can be supersymmetrized in such a way that the ground state
is a singlet with RSOS label $\hlf$ only. In particular, the corresponding
adjacency condition between the ground state and the nearest kink
is required to be satisfied. The supersymmetric part of the ground
state is denoted by $\ket{B_{\hlf}}$. The action of the boundary
supersymmetry generators in this case on $\ket{B_{\hlf}}$ is \begin{equation}
\tilde{Q}^{(-)}\ket{B_{\hlf}}=e\gamma\ket{B_{\hlf}},\quad\tilde{Z}\ket{B_{\hlf}}=0,\quad e=\pm1,\label{gr  st rep}\end{equation}
 where \[
\gamma\in\RR,\quad\gamma<0,\qquad{\textrm{or}}\qquad\gamma\in i\RR,\quad\gamma/i<0,\]
 and \[
\tilde{Q}^{(+)}\ket{B_{\hlf}}=\gamma\ket{B_{\hlf}},\quad\tilde{Z}\ket{B_{\hlf}}=0,\]
 where \[
\gamma\in\RR,\qquad{\textrm{or}}\qquad\gamma\in i\RR.\]
 The superscript of $\tilde{Q}$ stands for the $(+)$ and $(-)$
cases. $e\gamma$ or $\gamma$ is a parameter of the model to be described
and is expected to be expressible in terms of the parameters of the
Lagrangian density. The reason for writing this parameter in the form
$e\gamma$ in the $(-)$ case will become clear in \ref{sec 3.5}
when the ground state kink reflection amplitudes are discussed. Two
(symmetric) cases correspond to the sign $e$. We remark that singlet
ground states with RSOS label $1$ or $0$ are also possible.

\subsection{Factorized scattering\label{szorzat}}

Let the supersymmetric factor of the amplitude of a general multi-particle
process be $A$. The rapidities of the particles are allowed to take
complex values. Assume that there are $m$ individual two-particle
scatterings, reflections, fusions or decays in the process, which
are labeled by $1,2,...,m$ in chronological order ($t_{1}<t_{2}<...<t_{m}$).
$A$ can be written as a product of $m$ tensors, which can also be
viewed as linear maps: \[
A=T_{m}T_{m-1}T_{m-2}...T_{k}...T_{1},\]
 where the multiplication is the multiplication of linear maps, i.e.
the contraction of the appropriate upper and lower indices. If at
the time $t_{k}$ the {}``event'' is the scattering of the $n_{k}$-th
and $(n_{k}+1)$-th particle and by this time the \textsl{intermediate}
state contains $m_{k}$ one-particle states, then \[
T_{k}=I_{1}\otimes I_{2}\otimes...\otimes I_{n_{k}-1}\otimes S_{n_{k},n_{k}+1}\otimes I_{n_{k}+2}\otimes...\otimes I_{m_{k}},\]
 where $S_{n_{k},n_{k}+1}$ is the two-particle S-matrix for the $n_{k}$-th
and $(n_{k}+1)$-th particle. The $I_{l}$ is the identity tensor
for the SUSY indices of the $l$-th particle or for the SUSY indices
of the boundary state. Similar formulae can be written if the {}``event''
at $t_{k}$ is a reflection, fusion or decay.

The factorized form above has (among others) the following simple
implications: if $S_{n_{k},n_{k}+1}$ (or the reflection amplitude,
fusing or decay tensor) is bijective as a linear map, then $T_{k}$
is also bijective. If all the $T_{1},...T_{m}$-s are bijective, then
so is $A$. The $T_{k}$-s are equivariant.

\subsection{Supersymmetric reflection factors on the ground state boundary\label{sec 3.5}}

We consider one-particle kink reflection factors on ground state boundary
first. As the left and right RSOS labels should be conserved, $\{ R_{K}\}_{K_{1\hlf}}^{K_{0\hlf}}(\theta)=\{ R_{K}\}_{K_{0\hlf}}^{K_{1\hlf}}(\theta)=0$
must hold, i.e. $R_{K}$ is diagonal ($R_{K}$ denotes the kink reflection
factor). The general solution of the boundary Yang-Baxter equation,
unitarity condition and crossing equation without imposing supersymmetry
is \cite{key-10,key-5} \[
\{ R_{K}\}_{K_{0\hlf}}^{K_{0\hlf}}(\theta)=(1+A\sinh(\theta/2))M(\theta),\qquad\{ R_{K}\}_{K_{1\hlf}}^{K_{1\hlf}}(\theta)=(1-A\sinh(\theta/2))M(\theta),\]
 where $M(\theta)$ is restricted by unitarity and crossing symmetry.
After imposing the boundary supersymmetry condition one finds that
in the $(+)$ case \cite{key-2,key-9}\[
\{ R_{K}^{(+)}\}_{K_{0\hlf}}^{K_{0\hlf}}(\theta)=\{ R_{K}^{(+)}\}_{K_{1\hlf}}^{K_{1\hlf}}(\theta)=2^{-\theta/(\pi i)}P(\theta),\qquad A=0.\]
 ($P(\theta)$ can be found in the Appendix.) In the $(-)$ case there
are distinct solutions for a given $\gamma$ corresponding to the
sign $e$:\begin{equation}
\{ R_{K,e}^{(-)}\}_{K_{0\hlf}}^{K_{0\hlf}}(\theta)=(\cos\frac{\xi}{2}+ei\sinh\frac{\theta}{2})K(\theta-i\xi)K(i\pi-\theta-i\xi)2^{-\theta/(\pi i)}P(\theta),\label{refl1}\end{equation}
\begin{equation}
\{ R_{K,e}^{(-)}\}_{K_{1\hlf}}^{K_{1\hlf}}(\theta)=(\cos\frac{\xi}{2}-ei\sinh\frac{\theta}{2})K(\theta-i\xi)K(i\pi-\theta-i\xi)2^{-\theta/(\pi i)}P(\theta),\label{refl2}\end{equation}
 where $\gamma=-2\sqrt{M}\cos\frac{\xi}{2}$ and $0\leq\xi\leq\pi$
or $Re(\xi)=0$ or $Re(\xi)=\pi$, $M$ is the kink mass. (\ref{refl1})
and (\ref{refl2}) are invariant under $\xi\leftrightarrow-\xi$.
These reflection amplitudes are minimal, they have no poles and zeroes
in the physical strip. The sign $e$ seems to correspond here to the
$0$ and $1$ RSOS vacua. It should be noted that $R_{K}^{(+)}(\theta)$
is independent of $\gamma$ and $e$. Furthermore, as symmetry under
$\tilde{\Gamma}$ requires $R_{K_{0\hlf}}^{K_{0\hlf}}(\theta)=R_{K_{1\hlf}}^{K_{1\hlf}}(\theta)$,
the $R_{K}^{(+)}(\theta)$ is automatically $\tilde{\Gamma}$-symmetric,
although this is not required a priori. On the other hand, $R_{K,e}^{(-)}(\theta)$
are not $\tilde{\Gamma}$-symmetric. However, $\{ R_{K,e}^{(-)}\}_{K_{0\hlf}}^{K_{0\hlf}}(\theta)=-\{ R_{K,e}^{(-)}\}_{K_{1\hlf}}^{K_{1\hlf}}(\theta)$
if $\gamma=0$ ($A\rightarrow\infty$). We also remark that $\{ R_{K}\}_{K_{1\hlf}}^{K_{1\hlf}}(\theta)/\{ R_{K}\}_{K_{0\hlf}}^{K_{0\hlf}}(\theta)$
is determined by the supersymmetry condition, i.e. if we impose the
condition of invariance under supersymmetry, then we do not need to
solve the Yang-Baxter equation.

We determined the general solution of the Yang-Baxter equation for
the boundary supersymmetric particle reflection (on the ground state
boundary). We imposed the supersymmetry condition first, which is
$Q(-\theta)R(\theta)=R(\theta)Q(\theta),$ where $Q(\theta)$ is $\tilde{Q}$
on the states of the form $\ket{A(\theta)B_{\hlf}}$, and $R(\theta)$
is the particle reflection amplitude on ground state boundary. The
resulting forms of the reflection amplitude in the $(+)$ and $(-)$
cases are {\small \[
R_{P}^{(+)}(\theta)=Z^{(+)}(\theta)\frac{1}{\sqrt{m}}\left(\begin{array}{cc}
(X^{(+)}(\theta)+e\gamma Y^{(+)}(\theta))c(\frac{\theta}{2}-\frac{i\pi}{4}) & \sqrt{m}Y^{(+)}(\theta)c(\theta)\\
\sqrt{m}Y^{(+)}(\theta)c(\theta) & (X^{(+)}(\theta)-e\gamma Y^{(+)}(\theta))c(\frac{\theta}{2}+\frac{i\pi}{4})\end{array}\right),\]
 \[
R_{P}^{(-)}(\theta)=Z^{(-)}(\theta)\frac{1}{\sqrt{m}}\left(\begin{array}{cc}
(X^{(-)}(\theta)+e\gamma Y^{(-)}(\theta))c(\frac{\theta}{2}+\frac{i\pi}{4}) & i\sqrt{m}Y^{(-)}(\theta)c(\theta)\\
-i\sqrt{m}Y^{(-)}(\theta)c(\theta) & (X^{(-)}(\theta)-e\gamma Y^{(-)}(\theta))c(\frac{\theta}{2}-\frac{i\pi}{4})\end{array}\right),\]
} where $c$ stands for $\cosh$ and $X$, $Y$ and $Z$ are functions
not determined by supersymmetry. Now two cases can be distinguished
depending on whether $\tilde{\Gamma}$ is a symmetry or not: in the
first case, which is the $\tilde{\Gamma}$-symmetric case, $Y(\theta)\equiv0$,
$X(\theta)$ can be absorbed into the prefactor, and the structure
of the reflection amplitude is completely determined and does not
contain free parameters:\[
R_{P1}^{(\pm)}(\theta)=\frac{1}{\sqrt{m}}ZX^{(\pm)}(\theta)\left(\begin{array}{cc}
\cosh(\frac{\theta}{2}\mp\frac{i\pi}{4}) & 0\\
0 & \cosh(\frac{\theta}{2}\pm\frac{i\pi}{4})\end{array}\right).\]
 This case is discussed in \cite{key-4}, the explicit form of $ZX^{(\pm)}$
can be found in the Appendix, see also \cite{key-4,key-11,key-2}.
It can be checked that the boundary Yang-Baxter equation for incoming
particles of arbitrary masses is satisfied by this reflection amplitude.
$R_{P1}^{(\pm)}(\theta)$ can also be obtained from $R_{K}^{(+)}(\theta)$
and $R_{K}^{(-)}(\theta)$ at $\gamma=0$ by bootstrap \cite{key-2,key-11}.

In the second case, when $\tilde{\Gamma}$ is not conserved, $Y(\theta)$
is not identically zero, and it can be absorbed into the prefactor,
so one free function $y^{(\pm)}(\theta)=X^{(\pm)}(\theta)/Y^{(\pm)}(\theta)$
remains in the reflection amplitude, which is to be determined by
the boundary Yang-Baxter equation. To obtain $y^{(\pm)}(\theta)$
we solved the Yang-Baxter equation first in the case when the conserved
quantum numbers have the same values for the two incoming particles.
Although the boundary Yang-Baxter equation is quadratic in general,
in this case it is inhomogeneous linear in the variables $y^{(\pm)}(\theta_{1})$
and $y^{(\pm)}(\theta_{2}).$ The coefficient of the quadratic term
$y^{(\pm)}(\theta_{1})y^{(\pm)}(\theta_{2})$ vanishes precisely because
$R_{P1}^{(\pm)}(\theta)$ (!) satisfies the Yang-Baxter equation.
The Yang Baxter equation consists of 16 scalar equations in our case.
Some of them are trivial (0=0), and the remaining $n$ equations are
of the form\[
a_{q}^{(\pm)}(\theta_{1},\theta_{2})y^{(\pm)}(\theta_{1})+b_{q}^{(\pm)}(\theta_{1},\theta_{2})y^{(\pm)}(\theta_{2})+c_{q}^{(\pm)}(\theta_{1},\theta_{2})=0,\quad q=1..n.\]
 It is possible to choose two inequivalent equations from this set.
Two such equations can be solved for the numbers $y^{(\pm)}(\theta_{1})$
and $y^{(\pm)}(\theta_{2})$. The solution turns out to be of the
form $y^{(\pm)}(\theta_{1})=g^{(\pm)}(\theta_{1}),$ $y^{(\pm)}(\theta_{2})=g^{(\pm)}(\theta_{2})$
(for general coefficients $d_{q_{1}},e_{q_{1}},f_{q_{1}},d_{q_{2}},e_{q_{2}},f_{q_{2}}$
it would be of the form $y(\theta_{1})=g_{1}(\theta_{1},\theta_{2}),$
$y(\theta_{2})=g_{2}(\theta_{1},\theta_{2})$, which does not define
a function $y(\theta)$), where $g^{(\pm)}$ is a function that depends
also on $m$, $\alpha$, $\gamma$, but has no other parameters. Consequently,
the reflection amplitude depends on the conserved quantum numbers
through these parameters only. We checked that the solution obtained
in this way satisfies the other $n-2$ equations as well. In the next
step we checked that the solutions $R_{P2,e}^{(\pm)}(\theta)$ satisfy
the Yang-Baxter equation for incoming particles of different masses
as well. The two functions $y^{(+)}(\theta)$ and $y^{(-)}(\theta)$
have very similar form.

The solutions that we obtained can be brought to the following form:
\[
\{ R_{P2,e}^{(\pm)}\}_{b}^{b}(\theta)=A_{+}^{(\pm)}(\theta),\quad\{ R_{P2,e}^{(\pm)}\}_{f}^{f}(\theta)=A_{-}^{(\pm)}(\theta),\]
 \[
\{ R_{P2,e}^{(\pm)}\}_{b}^{f}(\theta)=\pm B^{(\pm)}(\theta),\quad\{ R_{P2,e}^{(\pm)}\}_{f}^{b}(\theta)=B^{(\pm)}(\theta),\]
 \[
A_{\pm}^{(-)}(\theta)=\tilde{Z}^{(-)}(\theta)\left\{ \cosh(\frac{\theta}{2})\left(\frac{\gamma^{2}}{4M}-\left[\sin^{2}(\frac{\rho}{4})+\sinh^{2}(\frac{\theta}{2})\right]\right)\right.\]
 \[
\left.\mp i\sinh(\frac{\theta}{2})\left(\frac{\gamma^{2}}{4M}+\left[\sin^{2}(\frac{\rho}{4})+\sinh^{2}(\frac{\theta}{2})\right]\right)\right\} ,\]
 \[
A_{\pm}^{(+)}(\theta)=\tilde{Z}^{(+)}(\theta)\left\{ -i\cosh(\frac{\theta}{2})\left(\frac{\gamma^{2}}{4M}-\left[\sin^{2}(\frac{\rho}{4})-\cosh^{2}(\frac{\theta}{2})\right]\right)\right.\]
 \[
\left.\pm\sinh(\frac{\theta}{2})\left(\frac{\gamma^{2}}{4M}+\left[\sin^{2}(\frac{\rho}{4})-\cosh^{2}(\frac{\theta}{2})\right]\right)\right\} \]
 \[
B^{(\pm)}(\theta)=\tilde{Z}^{(\pm)}(\theta)\frac{e\gamma}{2\sqrt{M}}\sqrt{\cos(\rho/2)}\sinh(\theta),\]
 where\begin{equation}
m=2M\cos(\frac{\rho}{2}),\quad\frac{\rho}{2}=\frac{\pi}{2}-u,\quad\alpha=-\frac{1}{2M},\label{def   rho}\end{equation}
 $0\leq\rho<\pi$, $e=\pm1$ in the $(-)$ case and $e=1$ in the
$(+)$ case. Note that $R_{P2,e}^{(\pm)}$ depends on two parameters:
$\gamma^{2}/M$ and $\rho$ only. $R_{P2,e}^{(-)}$ has the same structure
as the particle reflection amplitude obtained in \cite{key-2} for
the case of the boundary supersymmetric sine-Gordon model from the
kink reflection amplitude by bootstrap. Consequently, there is no
need now to solve the crossing and unitarity equations for $\tilde{Z}^{(-)}(\theta)$,
we take it from \cite{key-2}. We determined $\tilde{Z}^{(+)}(\theta)$
using the unitarity and crossing equations and exploiting the fact
that these equations take a similar form for $\tilde{Z}^{(-)}(\theta)$.
Explicit formulae for these prefactors can be found in the Appendix.

In the $(+)$ case we introduce the parameter $\xi$ in the following
way: $\gamma=-2\sqrt{M}i\sin(\xi/2)$, $-\pi\leq\xi\leq\pi$ or $Re(\xi)=0$
or $Re(\xi)=\pm\pi$.

To summarize, we have the solutions $R_{P2,e}^{(\pm)}$, which depend
on $\gamma$ and $e$ and are not symmetric with respect to $\tilde{\Gamma}$,
and we also have $R_{P1}^{(\pm)}$, which are independent of $\gamma$,
$e$ and are symmetric with respect to $\tilde{\Gamma}$. $R_{P1}^{(\pm)}$
and $R_{P2,e}^{(\pm)}$ do not satisfy the Yang-Baxter equation together.
It is also important to note that $\lim_{\gamma\rightarrow0}R_{P2,e}^{(\pm)}=R_{P1}^{(\pm)}$.
The poles and degeneracies of these reflection amplitudes will be
discussed in section \ref{ssy bndry bootstrap}.

The same set of kink and particle reflection factors can be obtained
by solving the Yang-Baxter equations without imposing the boundary
supersymmetry condition \cite{key-10,key-11,key-5}. The supersymmetry
condition relates the parameters of the reflection factors obtained
in this way to the parameters of the representations of the supersymmetry
algebra. The results described above show that the task of solving
the Yang-Baxter equations is greatly simplified if one imposes the
supersymmetry condition first.

\section{Supersymmetric boundary bootstrap\label{ssy bndry bootstrap}}

\begin{figure}
\begin{center}\includegraphics[%
  width=4cm,
  height=2.5cm]{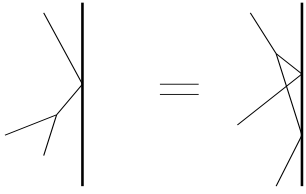}\end{center}

\caption{Boundary bootstrap equation I (BBE I)\label{fig bootstrap I}}
\end{figure}

The first boundary bootstrap equation (see fig. \ref{fig bootstrap I})
applied to the $kink+kink\rightarrow particle$ bulk fusion determines
reflection factors for particles on ground state boundary. They turn
out \cite{key-11,key-2} to be the same as those obtainable by solving
the boundary Yang-Baxter, crossing and unitarity equations. In terms
of the amplitudes $R_{K}^{(+)}+R_{K}^{(+)}\rightarrow R_{P1}^{(+)}$
and $R_{K,e}^{(-)}+R_{K,e}^{(-)}\rightarrow R_{P2,e}^{(-)}$ (with
appropriate values of the parameters). Similarly, it can be checked
that the first boundary bootstrap equation is also satisfied for the
$particle+particle\rightarrow particle$, $kink+particle\rightarrow kink$
fusions with the amplitudes $R_{P1}^{(\pm)}+R_{P1}^{(\pm)}\rightarrow R_{P1}^{(\pm)}$,
$R_{P2,e}^{(\pm)}+R_{P2,e}^{(\pm)}\rightarrow R_{P2,e}^{(\pm)}$ and
$R_{K}^{(+)}+R_{P1}^{(+)}\rightarrow R_{K}^{(+)}$, $R_{K,e}^{(-)}+R_{P2,e}^{(-)}\rightarrow R_{K,e}^{(-)}$
respectively. These relations are nontrivial, although it is clear
that they are satisfied up to CDD factors. It is remarkable that $R_{P2,e}^{(+)}$
cannot be obtained by bootstrap from kink reflection factors, while
the other particle reflection factors $R_{P1}^{(\pm)}$ and $R_{P2,e}^{(-)}$
can.

\subsection{Properties of the ground state reflection factors}

$R_{K}^{(+)}$ is bijective (of rank two) in the physical strip. $\{ R_{K,+1}^{(-)}\}_{K_{0\hlf}}^{K_{0\hlf}}(\theta)$
and $\{ R_{K,-1}^{(-)}\}_{K_{1\hlf}}^{K_{1\hlf}}(\theta)$ have a
zero at $\theta=i(\pi-\xi)$, so $R_{K,e}^{(-)}$ is of rank one at
this angle. This zero is in the physical strip if $\pi>\xi>\pi/2$,
any other zeroes of the kink amplitudes are outside the physical strip.

Consequently, the relations $R_{K}^{(+)}+R_{K}^{(+)}\rightarrow R_{P1}^{(+)}$,
and $R_{K,e}^{(-)}+R_{K,e}^{(-)}\rightarrow R_{P2,e}^{(-)}$ together
with the bijectivity of the kink-kink fusing tensor and $S_{K}$ imply
that $R_{P1}^{(+)}$ is of rank two (bijective) and has no poles in
the physical strip, and $R_{P2,e}^{(-)}(\theta)$ is also bijective
for generic values of $\theta,$ but it is of rank one if $\theta=i(\pi-\xi\pm\rho/2)$.
It is possible for these angles to be in the physical strip and on
the imaginary axis if $Im(\xi)=0$. In this case $(\pi-\xi-\rho/2)>-\pi/2$,
so if $(\pi-\xi-\rho/2)$ is negative, then there is a pole in the
physical strip at $i(\rho/2+\xi-\pi)$ because of unitarity. If $(\pi-\xi-\rho/2)>0,$
then $R_{P2,e}^{(-)}(\theta)$ has no poles and zeroes in the physical
strip, and within the physical strip it is of rank 1 if and only if
$\theta=i(\pi-\xi\pm\rho/2)$. We therefore impose the following condition
on $\xi$ (and also on $\gamma$) provided that $Im(\xi)=0$: \begin{equation}
\pi-\xi\geq\rho/2.\label{feltetel   0}\end{equation}
 $R_{P1}^{(-)}$ is also bijective and has no poles in the physical
strip. It can be verified by direct calculation that $R_{P2,e}^{(+)}(\theta)$
is degenerate (of rank two) at $\theta=i(\pi-\xi\pm\rho/2)$ and $\theta=i(\pi+\xi\pm\rho/2)$,
so the condition (\ref{feltetel 0}) reads in this case as\begin{equation}
\pi-|\xi|\geq\rho/2.\label{feltetel 0 vesszo}\end{equation}
 (Note that the relation between $\xi$ and $\gamma$ is different
in the $(+)$ and $(-)$ cases). If $\xi$ is not real, then all the
kink and particle reflection factors are nondegenerate on the imaginary
axis within the physical strip.

The relations above imply that it is consistent with the first boundary
bootstrap equation to describe the reflections on ground state boundary
of all kinks and particles in a theory by the minimal $R_{K}^{(+)}$,
$R_{K,e}^{(-)}$, $R_{P1}^{(\pm)}$, $R_{P2,e}^{(\pm)}$ solutions,
which have the property that they have neither poles nor (overall)
zeroes on the imaginary axis in the physical strip, so the pole structure
of the (full) ground state reflection factors is the same as in the
non-supersymmetric theory.

A particular boundary scattering theory is characterized apart from
the parameters of the bulk theory by the sign $(+)$ or $(-)$, and
also by the parameters $\gamma$ and $e$. In the $(+)$ case the
amplitudes $R_{K}^{(+)}$ and $R_{P1}^{(+)}$ are to be used if there
are both kinks and particles in the theory. If there are only particles
in the theory, then either $R_{P2,e}^{(+)}$ should be used for all
particles, or $R_{P1}^{(+)}$. In the $(-)$ case the amplitudes $R_{K,e}^{(-)}$
and $R_{P2,e}^{(-)}$ should be used as ground state reflection factors
if there are both kinks and particles in the theory. If there are
only particles in the theory, then either $R_{P2,e}^{(-)}$ or $R_{P1}^{(-)}$
should be used for all particles.

(\ref{feltetel 0}) and (\ref{feltetel 0 vesszo}) imply that if $R_{P2,e}^{(\pm)}$
describes the ground state reflections of the particles in a theory,
then \begin{equation}
\pi-|\xi|\geq\rho_{max}/2,\qquad\rho_{max}=\max_{i}(\rho_{i}),\label{feltetel 00}\end{equation}
 where $i$ runs over all particles, is necessary and sufficient for
all $R_{P2,e}^{(\pm)}$ to have no poles in the physical strip. We
remark, that the following CDD factors are consistent with the BBE
I: $E(\theta)=C(\theta-i\rho/2)C(\theta+i\rho/2)$ for particles with
mass $2M\cos(\rho/2)$, and $C(\theta)$ for kinks, where $C(\theta)$
is an arbitrary CDD factor. %

We introduce the following notation for the supersymmetric part of
boundary bound states: $\ket{\nu_{1},\nu_{2},...,\nu_{k},B_{\hlf}}$
denotes a multiplet (subspace) of states, where $\nu_{1}$, $\nu_{2}$,...,$\nu_{k}$
are the fusing angles in the successive steps of the creation of the
multiplet of states. The incoming bulk particle multiplet in the $n$th
step of the creation is $p_{n}(i\nu_{n})$. It is either a kink or
a boson/fermion multiplet, and if it is a boson/fermion multiplet,
then also an angle $u_{n}$ is associated to it (see (\ref{u-k})).
We will sometimes use the letter $\mu$ instead of $\nu$ if the incoming
bulk particle multiplet is a kink multiplet. We usually use the term
{}``state'' for a multiplet (subspace) of states from now on. We
call two states $\ket{\nu_{1},\nu_{2},...,\nu_{k},B_{\hlf}}$ and
$\ket{\nu_{1}',\nu_{2}',...,\nu_{k'}',B_{\hlf}}$ equal if and only
if $k=k'$, $\nu_{i}=\nu_{i}'$, and the type (boson/fermion or kink)
and mass of $p_{i}$ is the same as that of $p_{i}'$ $\forall i=1...k$.
In other words, two states are regarded equal if they are generated
in the same way. The number $k_{1}+k_{2}/2$ will be called the level
of the boundary state, where $k_{1}$ is the number of particles and
$k_{2}$ is the number of kinks among the $k=k_{1}+k_{2}$ incoming
particles. The notation for several kinks or particles in the bulk
is $\ket{p(i\theta_{1}),...,p(i\theta_{n}),\nu_{1},...,\nu_{k},B_{\hlf}}$
if the boundary is in the state $\ket{\nu_{1},...,\nu_{k},B_{\hlf}}$.
$i\theta_{1}...i\theta_{n}$ are the rapidities and $k(i\theta)$
is also used instead of $p(i\theta)$ for kinks. The {}``$B_{\hlf}$''
will often be omitted.

Two states will be called equivalent if they can be mapped to each
other by a bijective supersymmetry-equivariant map so that their corresponding
reflection and scattering amplitudes on any other states are also
equal. Our aim is to characterize the supersymmetric parts of the
states up to equivalence.

We introduce the shorthand $p_{1}(i\theta_{1})Fp_{2}(i\theta_{2})$
for the statement {}``$p_{1}(i\theta_{1})$ and $p_{2}(i\theta_{2})$
satisfy the bulk fusing conditions described in section \ref{bulk}''.
For the converse of this statement we use the shorthand $p_{1}(i\theta_{1})nFp_{2}(i\theta_{2}).$

\subsection{Supersymmetric boundary states generated by particles\label{4.1} }

In this and subsequent sections the subscripts and superscripts introduced
for two-particle S-matrices and one-particle reflection matrices will
often be suppressed, but this should not cause any confusion. The
following discussion applies to all the cases described in the previous
section. In this section and in \ref{sec 4.2} we assume, that the
ground state kink or particle reflection factors used are nondegenerate
at the rapidities $i\nu_{1},...,i\nu_{k}$, $\ket{\nu_{1},...,\nu_{k}}$
being the state under consideration. The discussion of the case when
this condition is not satisfied is deferred until the end of section
\ref{4.3}.

The following arguments include diagrams of processes, states, etc.,
but it is important that they are to be understood to represent algebraic
objects, expressions and relations. It is not required that the particular
non-supersymmetric theory to be supersymmetrized contains the counterparts
of all the states (for example) that will be introduced.

\subsubsection{First level boundary states\label{4.1.1}}

\begin{figure}
\begin{center}\includegraphics[%
  width=0.20\textwidth]{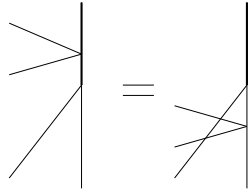}\end{center}

\caption{\label{fig bstr2} Boundary bootstrap equation II (BBE II)}
\end{figure}

A first level boundary bound state $\ket{\nu}$ has multiplicity 2,
and the corresponding fusing tensors are bijective. Reflection amplitudes
on a first level boundary state $\ket{\nu}$ are obtained by the second
bootstrap equation (see fig. \ref{fig bstr2}): \begin{equation}
R_{1}(\theta)[I\otimes g]=[I\otimes g][S_{P}(\theta+i\nu)\otimes I][I\otimes R(\theta)][S_{P}(\theta-i\nu)\otimes I],\label{bootstrap2}\end{equation}
 where $\theta$ is the rapidity of the reflecting particle $p_{1}$,
$R$ is a ground state particle reflection amplitude, $g$ is a boundary
fusing tensor, and $R_{1}$ is the reflection amplitude to be determined.

It is clear from the r.h.s. of eq. (\ref{bootstrap2}) that $R_{1}(\theta)$
has no poles if $\theta-i\nu$ (and also $\theta$) is on the imaginary
axis in the physical strip. This does not guarantee that $R_{1}(\theta)$
has no poles in the physical strip (on the imaginary axis) at all,
because $\theta-i\nu$ is not necessarily in the physical strip. $S_{P}(\theta-i\nu)$
on the r.h.s. has a pole at $\theta$ precisely if $p(i\nu)Fp_{1}(\theta)$,
where $\nu>\theta/i$. In this case $R_{1}$ also has a pole at $\theta$.
It can easily be checked that if $p(i\nu)Fp_{1}(\theta)$ then $S_{P}(\theta+i\nu)$
is bijective.

Consequently, assuming that $\ket{\nu}$ has multiplicity 2, $R_{1}(\theta)$
has no poles on the imaginary axis in the physical strip if and only
if $p(i\nu)Fp_{1}(\theta)$ is impossible for $\theta/i<\nu$, i.e.
\begin{equation}
\nu\leq u+u_{1},\label{feltetel  1}\end{equation}
 where $u$ is the angle associated (see (\ref{u-k})) to the particle
that generates the boundary state $\ket{\nu}$ and $u_{1}$ is the
angle associated to the reflecting particle $p_{1}(\theta)$. This
implies that in order for the supersymmetric reflection amplitudes
of all particles in a theory on the boundary state $\ket{\nu}$ to
have no poles on the imaginary axis in the physical strip it is necessary
and sufficient that\begin{equation}
\nu\leq u+u_{min},\label{feltetel 11}\end{equation}
 where $u_{min}=\min_{i}u_{i}$, $i$ runs over all particles in the
theory.

\subsubsection{The multiplicity of second level boundary states}

\begin{figure}
\begin{center}\includegraphics[%
  width=0.40\textwidth]{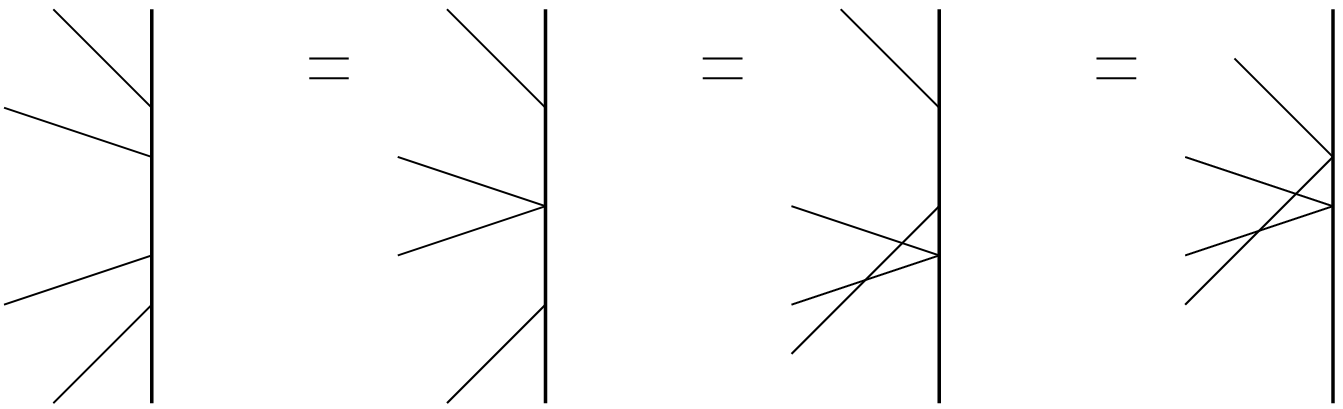}\end{center}

\caption{\label{fig eq1}}
\end{figure}

Let $\ket{\nu_{1},\nu_{2}}$ be a boundary bound state and assume
that $\ket{\nu_{1}}$ satisfies (\ref{feltetel 11}). We consider
first the case when $\nu_{2}>\nu_{1}$. In this case the amplitude
$A$ shown on the l.h.s of figure \ref{fig eq1} can be used to determine
the multiplicity and transformation properties of $\ket{\nu_{1},\nu_{2}}$.
The following equations hold in this case (see also fig. \ref{fig eq1}):\[
[I\otimes h_{1}]h_{2}g_{2}[I\otimes g_{1}]=[I\otimes h_{1}]R_{21}(i\nu_{2})[I\otimes g_{1}]=\]
 \[
=[I\otimes h_{1}][I\otimes g_{1}][S_{P}(i\nu_{2}+i\nu_{1})\otimes I][I\otimes R(i\nu_{2})][S_{P}(i\nu_{2}-i\nu_{1})I]=\]
\begin{equation}
=[I\otimes R(i\nu_{1})][S_{P}(i\nu_{1}+i\nu_{2})\otimes I][I\otimes R(i\nu_{2})][S_{P}(i\nu_{2}-i\nu_{1})\otimes I].\label{egyenlet1}\end{equation}
 If $S_{P}(i\nu_{1}+i\nu_{2})$ and $S_{P}(i\nu_{1}-i\nu_{2})$ are
bijective, then the multiplicity of the state $\ket{\nu_{1},\nu_{2}}$
is 4, and it transforms as the two-particle state $\ket{p(i\nu_{2}),p(i\nu_{1}),B_{\hlf}}$.

\begin{figure}
\begin{center}\subfigure[]{\includegraphics[width=0.20\textwidth]{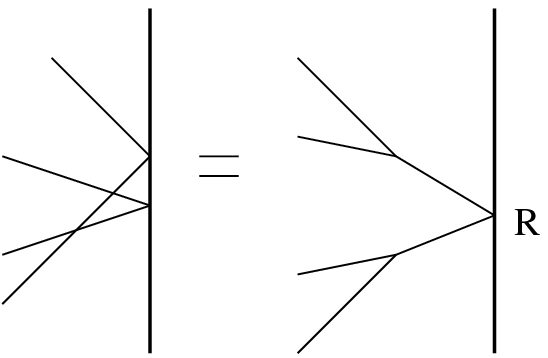}}\qquad\qquad\qquad \subfigure[]{\includegraphics[width=0.20\textwidth]{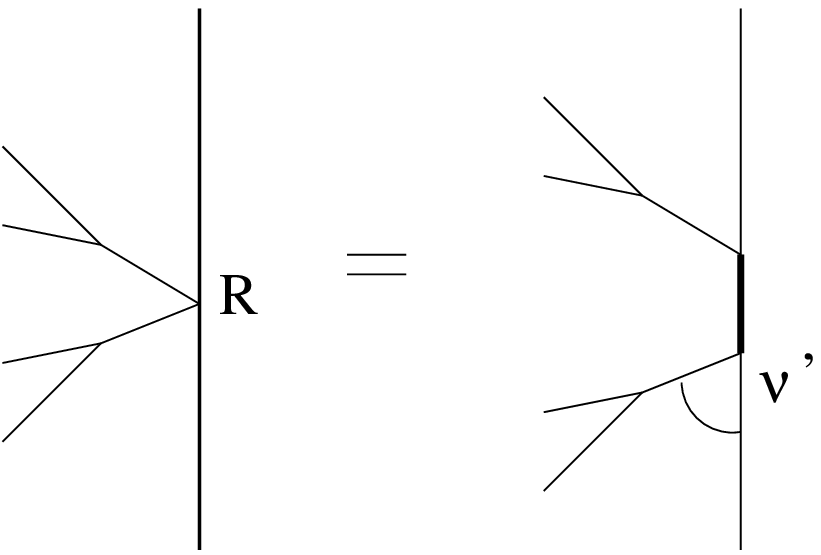}}\end{center}

\caption{\label{fig eq2} }
\end{figure}

If $S_{P}(i\nu_{1}-i\nu_{2})$ is degenerate, then the equation represented
in fig. \ref{fig eq2}a. holds.%

It is not hard to see, that the ground state reflection factor $R$
shown on the figure is bijective, so the equation implies that the
multiplicity of the state $\ket{\nu_{1},\nu_{2}}$ is 2 and it is
equivalent to $\ket{p(i\nu'),B_{\hlf}}$. Continuing the equation
shown on figure \ref{fig eq2}a. one can introduce a new first level
state $\ket{\nu'}$. This is shown on figure \ref{fig eq2}b. $\ket{\nu'}$
is equivalent to $\ket{\nu_{1},\nu_{2}}$. It is easy to verify that
$\nu'$ also satisfies condition (\ref{feltetel 11}).

\begin{figure}
\begin{center}\subfigure[]{\includegraphics[width=0.15\textwidth]{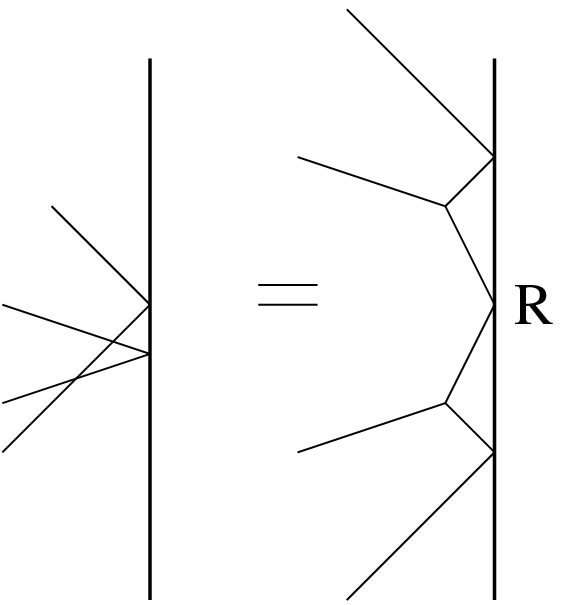}} \qquad\qquad\qquad \subfigure[]{\includegraphics[width=0.15\textwidth]{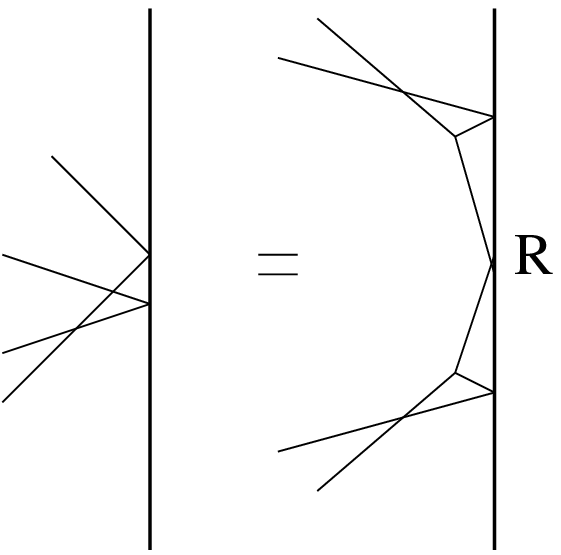}}\end{center}

\caption{\label{fig eq3}}
\end{figure}

If $S_{P}(i\nu_{1}-i\nu_{2})$ is bijective but $S_{P}(i\nu_{1}+i\nu_{2})$
is not, then the equation corresponding either to fig. \ref{fig eq3}a.
or to fig. \ref{fig eq3}b. holds, and it shows that the multiplicity
of the state $\ket{\nu_{1},\nu_{2}}$ is 2 and it is equivalent to
$\ket{p(i\nu'),B_{\hlf}}$. ($R$ is bijective.) Again a new $\ket{\nu''}$
first level state can be introduced, which is equivalent to $\ket{\nu_{1},\nu_{2}}$.
It is easy to verify that $\nu''$ also satisfies condition (\ref{feltetel 11}).

\begin{figure}
\begin{center}\includegraphics[%
  width=0.20\textwidth]{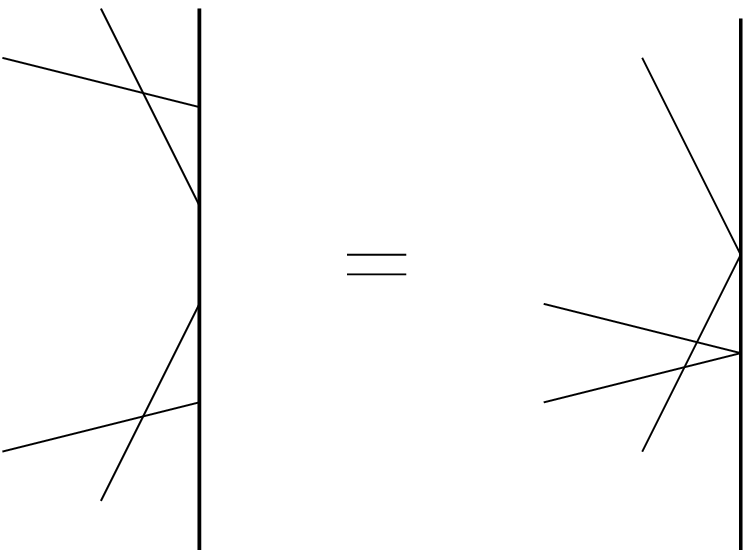}\end{center}

\caption{\label{fig ampl1}}
\end{figure}

In case of $\nu_{2}<\nu_{1}$ we consider the amplitude shown on the
left hand side of figure \ref{fig ampl1}. The condition (\ref{feltetel 11})
implies that $S_{P}(i\nu_{1}-i\nu_{2})$ is bijective, so this amplitude
is useful for determining the multiplicity of $\ket{\nu_{1},\nu_{2}}$.
An equation (which is analogous to (\ref{egyenlet1})) corresponding
to figure \ref{fig ampl1} is satisfied, and the same considerations
apply to the r.h.s. as in the $\nu_{1}<\nu_{2}$ case. It is clear
in particular that $\ket{\nu_{1},\nu_{2}}$ is equivalent to $\ket{\nu_{2},\nu_{1}}$
if these states have multiplicity 4, i.e. if $S_{P}(i|\nu_{1}-\nu_{2}|)$
and $S_{P}(i\nu_{1}+i\nu_{2})$ are bijective. %

\subsubsection{Scattering on second level boundary states}

Let $\ket{\nu_{1},\nu_{2}}$ be a second level boundary bound state
for which $S_{P}(i|\nu_{1}-\nu_{2}|)$ and $S_{P}(i\nu_{1}+i\nu_{2})$
are bijective. In any other case $\ket{\nu_{1},\nu_{2}}$ is equivalent
to some lower level state, and scattering on $\ket{\nu_{1},\nu_{2}}$
is equivalent to scattering on that lower level state. The same argument
as in subsection \ref{4.1.1} gives that condition (\ref{feltetel 11})
has to be satisfied also by $\nu_{2}$ in order for the reflection
amplitude on $\ket{\nu_{1},\nu_{2}}$ to have no poles on the imaginary
axis in the physical strip.

\subsubsection{Boundary states of arbitrary level\label{sec 4.1.4}}

The following results can be obtained by applying for higher level
states essentially the same steps and arguments as described in the
previous subsections. The strategy is the following: a multi-particle
reflection amplitude like the one on the l.h.s. of figure \ref{fig eq1}
or \ref{fig ampl1} should be considered at the values of rapidities
corresponding to the boundary bound state under consideration. The
corresponding diagram should be transformed into a diagram in which
no lines corresponding to boundary states occur (as on fig. \ref{fig eq1}).
Then the vertices at which the corresponding two-particle scattering
amplitudes are degenerate should be split, and the fusing and decaying
points should be moved sufficiently far away in the future and in
the past, while keeping the rest of the diagram fixed. If more than
one vertices can be split, then one of them should be chosen appropriately.
Finally, if there are no more vertices to split, then the diagram
is in a form that shows the transformation and scattering properties
of the boundary state manifestly.

Proposition 1: A general boundary bound state is equivalent to a boundary
state $\ket{\nu_{1},\nu_{2},...,\nu_{n}}$, which is such that $\nu_{i}$
satisfies condition (\ref{feltetel 11}) for $i=1..n$, and $p(i\nu_{i})nFp(i\nu_{j})$
and \\
 $p(i\nu_{i})nFp(-i\nu_{j})$ are satisfied for any $i,j=1..n$, $i\ne j$.
The multiplicity of this state is $2^{n}$ and the state is equivalent
to $\ket{p(i\nu_{1}),p(i\nu_{2}),...,p(i\nu_{n}),B_{\hlf}}$. The
reflection amplitudes on the state $\ket{\nu_{1},\nu_{2},...,\nu_{n}}$
have no poles on the imaginary axis in the physical strip. A boundary
state with these properties will be called irreducible. Two irreducible
states differing only in the order of the angles $\nu_{i}$ are equivalent.
Note that if $\ket{\nu_{1},\nu_{2},...,\nu_{n}}$ is an irreducible
state then so is $\ket{\nu_{1},\nu_{2},...,\nu_{k-1},\nu_{k+1},...,\nu_{n}}$
obtained by removing $\nu_{k}$.

Proposition 1 can be derived by induction on the level of boundary
states using Proposition 2.

Proposition 2: If an incoming particle of rapidity $i\nu_{n+1}$ fuses
with the irreducible state $\ket{\nu_{1},\nu_{2},...,\nu_{n}}$, then
three cases can be distinguished:

1st case: $p(i\nu_{n+1})nFp(i\nu_{k})$ and $p(i\nu_{n+1})nFp(-i\nu_{k})$
for $k=1..n$. The produced state is equivalent to the level $n+1$
state $\ket{\nu_{1},...,\nu_{n},\nu_{n+1}}$, which is irreducible.
$\nu_{n+1}$ has to satisfy (\ref{feltetel 11}) in this case in order
for the reflection factors on the produced state to have no poles
on the imaginary axis in the physical strip.

2nd case: $p(i\nu_{n+1})Fp(i\nu_{k})$ for one or more values of $k$,
$1\leq k\leq n$. Let $\nu_{k_{0}}$ be the greatest among the corresponding
$\nu_{k}$-s. The produced state is equivalent to \\
 $\ket{\nu_{1},...,\nu_{k_{0}-1},\nu_{k_{0}}',\nu_{k_{0}+1},...\nu_{n}}$,
where $\nu_{k_{0}}'$ can be computed by fusing $p(i\nu_{n+1})$ and
$p(i\nu_{k_{0}})$ (see fig. \ref{fig eq2}a. ). It is easy to verify
that $\nu_{k_{0}}'$ satisfies (\ref{feltetel 11}), so \\
 $\ket{\nu_{1},...,\nu_{k_{0}-1},\nu_{k_{0}}',\nu_{k_{0}+1},...\nu_{n}}$
has the property that $p(i\nu_{i})nFp(i\nu_{j})$ for $i,j=1..n$
and $\det(R(i\nu'_{k_{0}}))\ne0$, but $p(i\nu_{i})nFp(-i\nu_{j})$
is not necessarily true. If the latter property is also true for $\ket{\nu_{1},...,\nu_{k_{0}-1},\nu_{k_{0}}',\nu_{k_{0}+1},...\nu_{n}}$
then it is irreducible, otherwise the 3rd case applies to it with
$p(i\nu_{k_{0}}')$ being the incoming bulk particle and $\ket{\nu_{1},...,\nu_{k_{0}-1},\nu_{k_{0}+1},...\nu_{n},B_{\hlf}}$
the boundary state.

3rd case: $p(i\nu_{i})nFp(i\nu_{j})$ for $i,j=1..n$, but $p(i\nu_{n+1})Fp(-i\nu_{k})$
is true for certain values of $k$. Let $\nu_{k_{0}}$ be the smallest
of the corresponding angles. The produced state is equivalent to $\ket{\nu_{1},...,\nu_{k_{0}-1},\nu_{k_{0}}',\nu_{k_{0}+1},...\nu_{n}}$,
where $\nu_{k_{0}}'$ arises by fusing $p(i\nu_{n+1})$ and $p(-i\nu_{k_{0}})$
(see figs. \ref{fig eq3} a.,b.). It is easy to verify that $\nu_{k_{0}}'$
satisfies (\ref{feltetel 11}), so this state has the property that
$p(i\nu_{i})nFp(i\nu_{j})$ for $i,j=1..n$. $\det(R(i\nu_{k_{0}}'))\ne0$
is also true, but $p(i\nu_{i})nFp(-i\nu_{j})$ is not necessarily.
If the latter condition is also satisfied, then $\ket{\nu_{1},...,\nu_{k_{0}-1},\nu_{k_{0}}',\nu_{k_{0}+1},...\nu_{n}}$
is irreducible, otherwise the present case applies to it with $p(i\nu_{k_{0}}')$
being the incoming bulk particle and $\ket{\nu_{1},...,\nu_{k_{0}-1},\nu_{k_{0}+1},...\nu_{n}}$
the boundary state.

Even if applying the 2nd or the 3rd case does not result in an irreducible
state, the level of the involved boundary state decreases by one,
so an irreducible state is obtained in a finite number of steps.

The equivalences stated in Proposition 2 in the second and third cases
can be derived as follows: as in the previous subsections, one uses
a multi-particle reflection amplitude (see e.g. the r.h.s of fig.
\ref{fig eq1}) to determine the transformation properties of the
produced state. In the corresponding diagram the scattering vertex
of $p(i\nu_{n+1})$ and $p(i\nu_{k_{0}})$ should be split to a fusion
and a decay vertex first, and then the fusing and decaying points
should be moved far away in the past and future (using the Yang-Baxter
equations). This should be done in such a way that those reflections
which are not moved (with angles $i\nu_{k}$, $k\ne n+1,k_{0}$) should
eventually take place between the fusing and decaying points. If these
points are moved sufficiently far away, then because of the choice
of $k_{0}$ all two-particle scatterings before the fusing point and
after the decaying point are bijective, so only the part of the diagram
that lies between these points is relevant.

The statements above can be used to determine the multiplicities and
transformation properties of a boundary bound state in concretely
given models in a finite number of steps. $\ket{\nu_{1},\nu_{2},...,\nu_{k}}$
being the boundary state, the irreducible state equivalent to $\ket{\nu_{1},\nu_{2}}$
should be determined first, then using the result the state equivalent
to $\ket{\nu_{1},\nu_{2},\nu_{3}}$ should be determined, and so on.

\subsection{Boundary states generated by kinks and particles\label{sec 4.2}}

The statements and arguments in this subsection are similar to those
in \ref{sec 4.1.4}.

Proposition 1: A general boundary bound state is equivalent to a state
$\ket{\nu_{1},\nu_{2},...,\nu_{n}}$ which is such that $\nu_{i}$
satisfies condition (\ref{feltetel 11}) if $p_{i}(i\nu_{i})$ is
a particle, and $p_{i}(i\nu_{i})nFp_{j}(i\nu_{j})$ and $p_{i}(i\nu_{i})nFp_{j}(-i\nu_{j})$
are satisfied for any $i,j=1..n$, $i\ne j$. There is at most one
kink among the $p_{i}$-s. A state with these properties will be called
irreducible. The multiplicity of these states is $2^{n}$ and they
are equivalent to $\ket{p_{1}(i\nu_{1}),p_{2}(i\nu_{2}),...,p_{n}(i\nu_{n}),B_{\hlf}}$.
The reflection amplitudes on these states have no poles on the imaginary
axis in the physical strip. Two irreducible states differing only
in the order of the corresponding $p_{i}(i\nu_{i})$-s are equivalent.
Removing any angle from an irreducible state results in an irreducible
state.

Proposition 1 can be derived by induction on the level of boundary
states using Proposition 2.

Proposition 2: If an incoming particle of rapidity $i\nu_{n+1}$ fuses
with an irreducible boundary state, then the following four cases
can be distinguished:

1st case: the boundary state is $\ket{\nu_{1},\nu_{2},...,\nu_{n}}$,
where all the $p_{i}$-s ($i=1..n$) are particles and $p_{n+1}$
is also a particle. This situation is described in section \ref{4.1}.

2nd case: the boundary state is $\ket{\nu_{1},\nu_{2},...,\nu_{n}}$,
where all the $p_{i}$-s ($i=1..n$) are particles and $p_{n+1}$
is a kink. If $p(i\nu_{n+1})nFp(-i\nu_{k})$ is satisfied for $k=1..n$,
then the produced state $\ket{\nu_{1},...,\nu_{n},\nu_{n+1}}$ is
irreducible. If $p(i\nu_{n+1})Fp(-i\nu_{k})$ for some values of $k$,
then let $\nu_{k_{0}}$ be the smallest among the corresponding $\nu_{k}$-s.
The produced state is equivalent to the state $\ket{\nu_{1},...,\nu_{k_{0}-1},\nu_{k_{0}+1},...\nu_{n},\nu_{k_{0}}'}$,
where $\nu_{k_{0}}'$ can be computed by fusing $p(i\nu_{n+1})$ and
$p(i\nu_{k_{0}}),$ and $p(i\nu_{k_{0}}')$ is a kink. $\ket{\nu_{1},...,\nu_{k_{0}-1},\nu_{k_{0}+1},...\nu_{n},\nu_{k_{0}}'}$
is either irreducible or the present case applies to it with $p(i\nu_{k_{0}}')$
being the incoming bulk kink.

3rd case: the boundary state is $\ket{\nu_{1},\nu_{2},...,\mu_{m},...,\nu_{n}}$,
$p_{m}$ is a kink and $p_{n+1}$ is a particle. This is very similar
to the first case. It is possible that the produced state is equivalent
to the irreducible state $\ket{\nu_{1},\nu_{2},...,\mu_{m},...,\nu_{n},\nu_{n+1}}$.
In this case (\ref{feltetel 11}) must be satisfied for $\nu_{n+1}$.
The other possibility is that the produced state is equivalent to
the state $\ket{\nu_{1},...,\nu_{k_{0}-1},\nu_{k_{0}+1},...\nu_{n},\nu_{k_{0}}'}$,
where $p_{k_{0}}$ is either a kink or a particle. This state is either
irreducible or the present or the 2nd case applies to it with $p_{k_{0}}$
being the incoming particle.

4th case: The boundary state is $\ket{\nu_{1},\nu_{2},...,\mu_{m},...,\nu_{n}}$,
where $p_{m}$ is a kink and the incoming particle $p_{n+1}$ is also
a kink. In this case the vertex of the two kinks $p(i\mu_{m})$ and
$p(i\nu_{n+1})$ can be split and the fusing and decaying points should
be moved far away to the past and future. Thus the produced state
is equivalent to $\ket{\nu_{1},\nu_{2},...,\nu_{m}',...,\nu_{n}}$,
where $p(i\nu_{m}')$ is a particle, $\nu_{m}'$ is determined by
$\mu_{m}$ and $\nu_{n+1}$. Now the first case can be applied with
$p(i\nu_{m}')$ being the incoming bulk particle and $\ket{\nu_{1},\nu_{2},...,\nu_{m-1},\nu_{m+1},...,\nu_{n}}$
being the boundary state. It is easy to see that $\nu'$ satisfies
the condition (\ref{feltetel 11}) in a stronger form:\begin{equation}
\nu'\leq u'.\label{feltetel2}\end{equation}

The statements above can be used to determine the multiplicities and
transformation properties of a boundary bound state in concretely
given models in a finite number of steps (see also the end of section
\ref{sec 4.1.4}).

An important simple special case is the following: if $\ket{\mu_{1},\mu_{2},...,\mu_{n}}$
is such that all the $p_{i}$-s ($i=1..n$), are kinks and $\mu_{i}\ne\mu_{j}$
for all $i\ne j$, then the multiplicity of this state is the maximal
$2^{\left\lceil n/2\right\rceil }$. This follows immediately from
the bijectivity of $S_{K}$, but could also be derived using the propositions
above. The reflection amplitudes on such a state have no poles on
the imaginary axis in the physical strip.

We remark that irreducible states that are not equal in our terminology
can be equivalent.

In the next subsection we bring the results of the present and \ref{4.1}
subsections to a simpler form.

\subsection{Supersymmetric boundary states and fusing rules\label{4.3}}

To a supersymmetric particle (multiplet) $p(\theta)$ with mass $m=2M\cos(\rho/2)$
we assign the following (not ordered) set of rapidities: $L_{p(\theta)}=\{\theta-i\rho/2,\theta+i\rho/2\}$,
where it is not required that $\theta\pm i\rho/2$ be in the physical
strip. The elements of the set are the rapidities of those kinks which
fuse to the particle $p(\theta)$. $L_{p(\theta)}$ determines $p(\theta)$
uniquely. We assign the set $L_{k(\theta)}=\{\theta\}$ to a kink
$k(\theta)$. This notation allows us to summarize the bulk supersymmetric
fusing rules in a simple form. In terms of these sets the bulk fusion
$p_{1}+p_{2}\rightarrow p_{3}$ of two particles takes the form \begin{equation}
\{\theta_{1},\theta_{2}\}+\{\theta_{3},\theta_{1}\pm i\pi\}\rightarrow\{\theta_{2},\theta_{3}\},\label{fr1}\end{equation}
 where $L_{p_{1}}=\{\theta_{1},\theta_{2}\}$, $L_{p_{2}}=\{\theta_{3},\theta_{1}\pm i\pi\}$,
$L_{p_{3}}=\{\theta_{2},\theta_{3}\}$, and $\theta_{1},\theta_{2},\theta_{3}$
are appropriate complex rapidities. Similarly, the fusion of a kink
and a particle $k_{1}+p\rightarrow k_{2}$ takes the form \begin{equation}
\{\theta_{1}\}+\{\theta_{2},\theta_{1}\pm i\pi\}\rightarrow\{\theta_{2}\}.\label{fr2}\end{equation}
 A kink-kink fusion $k(\theta_{1})+k(\theta_{2})\rightarrow p$ takes
the form \begin{equation}
\{\theta_{1}\}+\{\theta_{2}\}\rightarrow\{\theta_{1},\theta_{2}\}.\label{fr3}\end{equation}
 The essential point is that in these fusions the set of rapidities
corresponding to the final state is obtained in the following way:
the disjoint union of the two sets of rapidities corresponding to
the fusing particles/kinks is formed and the pair of rapidities differing
by $\pm i\pi$ are deleted (if there is any such pair). We allow here
and further on that a set contains certain elements several times,
i.e. the elements of the sets we consider have multiplicity.

These rules are in accordance with the fact that $S_{K}$ is bijective
(of rank 6) in the physical strip but is of rank 3 at $\theta=i\pi$.
The fusing rules (\ref{fr1}), (\ref{fr2}), (\ref{fr3}) apply in
the same form if we do not perform the factorizations described in
section \ref{sec 2.3}, i.e. if two kinks fuse into a state with multiplicity
6, which we call {}``two-kink-state''. Correspondingly, the fusing
rules given in section \ref{sec 2.3} also apply in this case with
the minor modification that {}``two-kink-states'' should be substituted
for {}``supersymmetric particles''. In this case the fusion of two
kinks produces a {}``two-kink-state'' state with multiplicity 6
as described in sections \ref{sec 2.2} and \ref{sec 2.3}. It does
not seem to be useful to stress the decomposition (\ref{cg3}) because
of the adjacency conditions and relations following from the kink
adjacency conditions. Numerical (TCSA) calculations in finite volume
suggest \cite{key-19} that it is possible that the breathers of the
bulk supersymmetric sine-Gordon model transform as {}``two-kink-states''
rather than simple SUSY particles \cite{key-15}.

Let us assume that the particle reflections are described by the factors
$R_{P1}^{(\pm)}$ or $R_{P2,e}^{(-)}$, i.e. by the factors that can
be obtained by bootstrap from the kink reflection factors. The case
of $R_{P2,e}^{(+)}$ will be considered later.

To a general boundary state $v=\ket{\nu_{1},\nu_{2},...\nu_{n}}$
we assign the set $P_{v}$ the elements of which are the sets of rapidities
assigned to $p_{1},p_{2},...,p_{n}$ with the modification that all
the rapidities $i\theta$ ($\theta\in\RR$) are changed to $i|\theta|$.
Note that $0\leq|\theta|<\pi$. For example, $P_{v}=\{\{ i|\theta_{11}|,i|\theta_{12}|\},\{...\},...\}$
if $p_{1}$ is a particle and $L_{p_{1}(i\nu_{1})}=\{ i\theta_{11},i\theta_{12}\}$.
The disjoint union of the sets contained by $P_{v}$ is denoted by
$\cup P_{v}$. It is easily verified that $v$ is irreducible if and
only if $\cup P_{v}$ does not have two elements $i\theta_{1},i\theta_{2}$
for which $\theta_{1}+\theta_{2}=\pi$.

Let $v'$ denote the (not necessarily unique) irreducible state obtained
from the boundary state $v$ by applying the results of section \ref{4.1}
and \ref{sec 4.2}. It is straightforward to verify that the effect
of the steps (described in \ref{sec 4.1.4} and \ref{sec 4.2}) used
to obtain $v'$ is that pairs of rapidities in $\cup P_{v}$ with
the property $\theta_{1}+\theta_{2}=\pi$ are removed. Consequently,
$\cup P_{v'}$ is obtained from $\cup P_{v}$ by removing all such
pairs. We introduce the notation $L_{v}=\cup P_{v'}$.

Using BBE I it is easy to see that an irreducible state $w$ is equivalent
to the state $\tilde{w}=\ket{\theta_{1},\theta_{2},...,\theta_{n}}$,
where $p(i\theta_{i})$ is a kink for all $i$-s, $L_{w}=\cup P_{w}=\{ i\theta_{i}\,|\, i=1...n\}$,
and $0<\theta_{1}\leq\theta_{2}\leq...\leq\theta_{n}$. However, some
of the $\theta$-s can be outside the physical strip here. This state
has multiplicity $2^{\left\lceil n/2\right\rceil }$ and it is equivalent
to $\ket{k(i\theta_{1}),k(i\theta_{2}),...,k(i\theta_{n}),B_{\hlf}}$.

Now it is clear that two irreducible states $w_{1}$ and $w_{2}$
are equivalent if $L_{w_{1}}=L_{w_{2}}$. This condition is also necessary:
the bootstrap equations applied to $\tilde{w}$ (where $w$ is an
irreducible state) imply that the one-kink reflection amplitude $r(i\theta)$
on $w$ can also be expressed as a product containing the factors
$S_{K}(i(\theta\pm\theta_{i})$, $R_{K}(i\theta)$ and $\theta$-independent
numerical factors, where $i\theta$ is the rapidity of the reflecting
kink. Thus the sequence $\theta_{i}$, $i=1...n$ is uniquely determined
by the analytic properties of $r(i\theta)$, so if $w_{1}$ and $w_{2}$
are equivalent, then $L_{w_{1}}=L_{w_{2}}$ must hold.

The supersymmetric boundary fusion rules can be summarized as follows:
a supersymmetric boundary state $v$ is equivalent to the state $\widetilde{(v')}$.
Two supersymmetric boundary bound states $v$, $z$ are equivalent
if and only if $L_{w}=L_{z}$ (or equivalently $\widetilde{(v')}=\widetilde{(z')}$).
The reflection factors on $v$ have no poles and zeroes on the imaginary
axis in the physical strip if and only if \begin{equation}
\theta_{i}<\pi-\rho_{max}/2\quad\forall i=1...n.\label{feltetel 4}\end{equation}
 (See (\ref{feltetel 00}) for $\rho_{max}$, $L_{v}=\{ i\theta_{i}\,|\, i=1...n\}$)
Consider now the boundary fusion $p+v\rightarrow y$, where $p$ is
a bulk particle, $v$ and $y$ are boundary states. $L_{y}$ can be
obtained from $L_{p}$ and $L_{v}$ in the following way: the disjoint
union $L_{p}\cup L_{v}$ should be formed, the rapidities $i\theta\in L_{p}\cup L_{v}$
should be replaced by the values $i|\theta|$, and all the pairs $i\theta_{1},i\theta_{2}$
satisfying the condition $\theta_{1}+\theta_{2}=\pi$ should be removed.
This rule is analogous to the bulk fusion rules, but the amplitudes
$i\theta$ and $-i\theta$ ($\theta\in\RR$) are identified in this
case.

It is also possible that (\ref{feltetel 4}) is not satisfied for
certain bound states, but the poles of the (supersymmetric) reflection
factors on them are canceled by zeroes of the corresponding non-supersymmetric
reflection factor, so the poles of the supersymmetric reflection factors
do not introduce new bound states. The fusion rules above apply in
this case as well.

These boundary fusion rules can be applied unchanged if instead of
supersymmetric particles {}``two-kink-states'' are taken.

The statements above also apply to the case when particle reflections
on the ground state are described by the factor $R_{P2,e}^{(+)}$
with the following modification: the rapidities should not be altered
when $P_{v}$ is constructed, i.e. for example $P_{v}=\{\{ i\theta_{11},i\theta_{12}\},\{...\},...\}$.
Although a rapidity $\theta$ in $\cup P_{v}$ is not identified with
$-\theta$, $\cup P_{v_{1}}$ and $\cup P_{v_{2}}$ are regarded equivalent
if they can be obtained from one another by changing the sign of an
even number of rapidities. $L_{v}$ is obtained from $\cup P_{v}$
by removing all the pairs of amplitudes of the form $i\theta,-i\theta$,
taking into consideration that the sign of an even number of amplitudes
in $\cup P_{v}$ can be changed freely. If a change of signs gives
rise to pairs of the form $i\theta,-i\theta$, then this change should
be done, and the arising pairs should be removed. $w$ is equivalent
to $\ket{k(i\theta_{1}),k(i\theta_{2}),...,k(i\theta_{n}),B_{\hlf}}$,
where $L_{w}=\{ i\theta_{1},...,i\theta_{n}\}$. The states $v$ and
$w$ are equivalent if and only if $L_{v}$ can be transformed into
$L_{w}$ by changing the sign of an even number of rapidities. We
use the notation $L_{v}\sim L_{w}$ in this case. $L_{y}$ is obtained
from $L_{p}\cup L_{v}$ in the same way as $L_{v}$ from $\cup P_{v}$.

These modifications are due to the fact that $R_{P2,e}^{(+)}$ is
not obtained from a kink amplitude by bootstrap and the boundary Yang-Baxter
equation is not satisfied by $R_{P2,e}^{(+)}$ and $R_{K}^{(+)}$.

We consider now the case when kink rapidities $i(\pi-|\xi|)$ are
also allowed, and the reflections on the ground state are described
by the factors $R_{K,e}^{(-)}$, $R_{P2,e}^{(-)}$ or $R_{P2,e}^{(+)}$.
This is the case when the ground state reflection factors are allowed
to be degenerate at particular fusing rapidities (see also the first
paragraph of section \ref{4.1}). The description above is modified
in this case as follows: it is not allowed to change the sign of all
the rapidities $i(\pi-|\xi|)$ in a set in the $(+)$ case.

In the $(-)$ case, if $L_{v}$ contains $i(\pi-|\xi|)$, then the
state $v$ is equivalent to the state $\ket{k(i\theta_{1}),k(i\theta_{2}),...,k(i\theta_{l-1}),k(i\theta_{l+1}),...,k(i\theta_{n}),B_{0/1}}$,
where $i\theta_{1},...,i\theta_{n}$ are the elements of $L_{v}$
(they are not necessarily different), $\theta_{l}=\pi-|\xi|$, and
$\ket{B_{0/1}}=\ket{\pi-|\xi|}$. It is allowed that $|\theta_{j}|=\pi-|\xi|$
for several integers $j\ne l$. The multiplicity of $v$ is $2^{\left\lceil n/2\right\rceil -1}$.
$\ket{B_{0/1}}$ is a singlet state.

In the $(+)$ case $v$ is equivalent to \\
 $\ket{k(i\theta_{1}),k(i\theta_{2}),...,k(i\theta_{l-1}),k(i\theta_{l+1}),...,k(i\theta_{j-1}),k(i\theta_{j+1}),...,k(i\theta_{n}),B_{\hlf}'}$,
where $j$ is chosen arbitrarily and $\ket{B_{\hlf}'}=\ket{\nu'}$,
$p(i\nu')$ is the particle constituted by the kinks $k(i\theta_{l})$
and $k(i\theta_{j})$. $\ket{B_{\hlf}'}$ is a singlet state.

The rules and statements of the present subsection are sufficient
to determine the transformation and scattering properties of the boundary
bound states in concretely given models without the use of the results
described in \ref{sec 4.1.4} and \ref{sec 4.2}.

Further supersymmetric boundary theories may be obtained from known
ones by projection, i.e. by factoring out certain boundary bound states.
We do not discuss this possibility in detail.

It is important to note that we have not taken into consideration
the {}``statistics'' of the various bulk particles at all, i.e.
we have assumed {}``free'' statistics. The correct statistical
properties can be achieved by  projection. An example of this 
can be found in section \ref{sec free part}.  

Let us consider now the situation that a boundary state in the non-supersymmetric
theory can be created in more than one way, i.e. the states $v$ and
$v'$ are equivalent (and so regarded to be the same) in the non-supersymmetric
theory. Such an equivalence is usually established by transforming
certain diagrams representing amplitudes into each other, where the
allowed transformations include shifting lines using bootstrap and
Yang-Baxter equations and splitting or fusing vertices. These transformations
are similar to those done above, and we expect that they can usually
be used to establish the equivalence of the corresponding supersymmetric
states as well. Therefore we expect that the supersymmetric states
$Sv$ and $Sv'$ corresponding to $v$ and $v'$ are also equivalent,
i.e. $L_{Sv}=L_{Sv'}$ (or $L_{Sv}\sim L_{Sv'}$). The equivalence
of such states should be checked in specific models, but we expect
that the requirement that such states should be equivalent is usually
satisfied and does dot place constraints on the boundary fusing rules
of the non-supersymmetric theory. The only really restrictive condition
on the applicability of the Ansatz described in the previous sections
is thus (\ref{feltetel 4}).

\section{Examples}

The following statement illustrates the phenomenon also encountered
in the bulk scattering theory \cite{key-3,key-1} that prescribing
the representations in which the boundary states transform places
very restrictive conditions on the boundary fusing rules. Let $P$
be a supersymmetric particle with parameters $m=2M\sin(u)$ (where
$M>0$, $0<u\leq\pi/2$). Let the boundary state $w=\ket{\nu_{1},\nu_{2},...,\nu_{n}}$
be generated by this particle in $n$ steps and assume that the ground
state particle reflection factor is nondegenerate at $i\nu_{i}$,
$i=1..n$. In this case $\ket{\nu_{1}}$, $\ket{\nu_{1},\nu_{2}}$,...,$w$
have multiplicity 2 (and are equivalent to first level states) if
and only if $\nu_{k}-\nu_{k-1}=2u$ $\forall k=2...n$.

For the sake of simplicity, we do not consider the case when particle
reflections on the ground state are described by $R_{P2,e}^{(+)}$
in the next subsections.

\subsection{Boundary sine-Gordon model}

The particle spectrum of the bulk sine-Gordon theory (SG) contains
a soliton ($s$) and an antisoliton ($\bar{s}$) of mass $M$ and
the breathers $B_{n}$ of mass $m_{n}=2M\sin(u_{n})$, where $u_{n}=\pi n/(2\lambda)$,
$n=1,...,[\lambda]$, $\lambda$ and $M$ are parameters of the model.
The breathers are self-conjugate, and the conjugate of $s$ is $\bar{s}$.
The fusing rules (given in the form process, (fusing angle)) are the
following: $s+\bar{s}\rightarrow B_{n}$, $(\pi-2u_{n})$; $B_{n}+B_{m}\rightarrow B_{n+m}$,
$(u_{n}+u_{m})$ provided $n+m\leq[\lambda]$; and the crossed versions
of these rules. They are consistent with associating kink representations
to $s$ and $\bar{s}$ and particle or {}``two-kink-state'' representations
with $\alpha=-1/(2M)$ to the breathers. The sets corresponding to
the supersymmetric parts of $s$, $\bar{s}$, and $B_{n}$ are $L_{s}=\{\theta\}$,
$L_{\bar{s}}=\{\theta\}$, $L_{B_{n}}=\{\theta+i(\pi/2-u_{n}),\theta-i(\pi/2-u_{n})\}$.

The boundary SG model (BSG) as defined in \cite{key-17} (see also
\cite{key-16}) has the boundary spectrum containing the states $\ket{n_{1},n_{2},...,n_{k}}$,
where $n_{1},n_{2},...,n_{k}$ are nonnegative integers satisfying
the condition $\pi/2\geq\nu_{n_{1}}>w_{n_{2}}>\nu_{n_{3}}>...\geq0$,
where $\nu_{n}=\eta/\lambda-u_{2n+1}$, $w_{n}=\pi-\eta/\lambda-u_{2n-1}$,
and $0<\eta\leq\frac{\pi}{2}(\lambda+1)$ is a boundary parameter.
The fusing rules \cite{key-16,key-24} are listed in the following
table: \vspace{0.5cm}

\begin{center}\begin{tabular}{|c|c|c|c|}
\hline 
Initial state&
 particle&
 rapidity&
 final state\tabularnewline
\hline
$|n_{1},...,n_{2k}\rangle$&
 $s,\bar{s}$&
 $i\nu_{n}$&
 $|n_{1},...,n_{2k},n\rangle$\tabularnewline
\hline
$|n_{1},...,n_{2k-1}\rangle$&
 $s,\bar{s}$&
 $iw_{n}$&
 $|n_{1},...,n_{2k-1},n\rangle$\tabularnewline
\hline
$|n_{1},...,n_{2k},n_{2k+1},...\rangle$&
 $B^{n}$&
 $i\frac{1}{2}(\nu_{l}-w_{n-l})$&
 $|n_{1},...,n_{2k},l,n-l,n_{2k+1},...\rangle$\tabularnewline
\hline
$|n_{1},...,n_{2k-1},n_{2k},...\rangle$&
 $B^{n}$&
 $i\frac{1}{2}(w_{l}-\nu_{n-l})$&
 $|n_{1},...,n_{2k-1},l,n-l,n_{2k},...\rangle$\tabularnewline
\hline
$|n_{1},...,n_{2k},...\rangle$&
 $B^{n}$&
 $i\frac{1}{2}(\nu_{-n_{2k}}-w_{n+n_{2k}})$&
 $|n_{1},...,n_{2k}+n,...\rangle$\tabularnewline
\hline
$|n_{1},...,n_{2k-1},...\rangle$&
 $B^{n}$&
 $i\frac{1}{2}(w_{-n_{2k-1}}-\nu_{n+n_{2k-1}})$&
 $|n_{1},...,n_{2k-1}+n,...\rangle$ \tabularnewline
\hline
\end{tabular}\end{center}

\vspace {0.5cm}

The first two lines show that the whole boundary spectrum can be generated
by kinks. Correspondingly, we associate to the BSG state $\ket{n_{1},n_{2},...,n_{k}}$
the supersymmetric part $\ket{\nu_{n_{1}},w_{n_{2}},\nu_{n_{3}},...B_{\hlf}}$
(using the notation introduced earlier), where $p(i\nu_{n_{1}})$,
$p(iw_{n_{2}})$... are kinks. Now we have to verify if the fusion
rules given in the 3-6th lines are also valid for these supersymmetric
parts. This is easily done by using the rules given in section \ref{4.3}.
Let us consider the 3rd line first. Let $v=\ket{\nu_{n_{1}},...,w_{n_{2k}},\nu_{n_{2k+1}},...,B_{\hlf}}$,
$p=p(i\hlf(\nu_{l}-w_{n-l}))$ with $u_{n}$, $w_{2k}>\nu_{l}>w_{n-l}>\nu_{2k+1}$,
and $p+v\rightarrow y$. In this case $L_{p}=\{ i\nu_{l},-iw_{n-l}\}$,
so $L_{y}=\{ i\nu_{n_{1}},\dots,iw_{n_{2k}},i\nu_{n_{2k+1}},\dots,i\nu_{l},iw_{n-l}\}$.
The 4th line is similar. Turning to the 5th line, let $v=\ket{\nu_{n_{1}},...,w_{n_{2k}},...,B_{\hlf}}$,
$p=p(i\hlf(\nu_{-n_{2k}}-w_{n+n_{2k}}))$ with $u_{n}$, and $p+v\rightarrow y$.
Now $L_{p}=\{ i\nu_{-n_{2k}},-iw_{n+n_{2k}}\}$, and because of $\nu_{-n_{2k}}+w_{n_{2k}}=\pi$
we have $L_{y}=\{ i\nu_{n_{1}},...,iw_{n+n_{2k}},...\}$. The 6th
line is similar to the 5th line.

Condition (\ref{feltetel 4}) is clearly satisfied for all boundary
states, so the supersymmetric factors of the reflection amplitudes
have no poles on the imaginary axis in the physical strip.

We remark that the one-particle reflection amplitudes and two-particle
S-matrices of the BSG model contain two bulk and two boundary parameters.
The relation between these parameters and the parameter $\gamma$
(see (\ref{gr st rep})) in the $(-)$ case is not yet known, so it
cannot be decided whether any of the angles $\nu_{n}$ and $w_{n}$
coincides with $\pi-\xi$ or not. The transformation and scattering
properties of the boundary states are modified by such a coincidence.

Because of the solitons the factor $R_{P2,e}^{(+)}$ cannot be used
to describe the reflections of breathers on the ground state boundary.
This factor can be used, however, if one restricts to the breather
sector of the model.

\subsection{Boundary sinh-Gordon model\label{sec BShG}}

The boundary sinh-Gordon (BShG) model is obtained from the BSG model
by analytic continuation in the bulk coupling constant $\beta$, where
$\beta$ is related to $\lambda$ through $\lambda=\frac{8\pi-\beta^{2}}{\beta^{2}}$.
In ShG $\beta=i\hat{\beta}$, where $\hat{\beta}\in\RR$, so $\lambda=-\frac{8\pi+\hat{\beta}^{2}}{\hat{\beta}^{2}}<-1$.
The particle spectrum of the ShG model contains only one self-conjugate
particle $P$ with a two-particle scattering amplitude and reflection
amplitude that can also be obtained from the corresponding amplitudes
of the first SG breather ($B_{1}$) by the analytic continuation $\beta=i\hat{\beta}$.
BShG also contains a series of boundary states $b_{n}$ \cite{key-6}
corresponding to the BSG states $\ket{\nu_{0},w_{n}}$, $n=1...$
which are precisely those states that can be generated using $B_{1}$
only. $b_{0}$ is the ground state. The bulk fusing rules are the
following: \vspace{0.5cm}

\begin{center}\begin{tabular}{|c|c|c|c|}
\hline 
Initial state&
 particle&
 rapidity&
 final state\tabularnewline
\hline
$b_{n}$&
 $P$&
 $i(\frac{\eta}{\lambda}-\frac{\pi}{2}+\frac{\pi}{\lambda}n)$&
 $b_{n+1}$ \tabularnewline
\hline
\end{tabular}\end{center}

\vspace{0.5cm}

The supersymmetric BShG model is also obtained by the analytic continuation
above. This implies that in the formula $m=2M\sin(u)$ for the particle
mass $M<0$ and $u<0$, i.e. the values of these parameters are not
in the range that is considered in our paper. A consequence of this,
for example, is that the supersymmetric factor of the two-particle
scattering amplitude has a pole in the physical strip (canceled by
a zero of the non-supersymmetric factor) \cite{key-12}. However,
the supersymmetric parts of the states $b_{n}$ and the corresponding
reflection factors can be obtained by the same steps as those of $\ket{\nu_{0},w_{n}}$
but with $M<0$, $u<0$, so the diagrams cannot be drawn as certain
angles take non-physical values. Thus the states $b_{n}$ have multiplicity
two in the supersymmetric BShG model if the value of $\xi$ is generic.
The situation is similar if the ground state reflection factor is
$R_{P2,e}^{(+)}$.

\subsection{Free particle on the half line\label{sec free part}}

The supersymmetric factors of the scattering and ground state reflection
amplitudes of this model can be obtained by taking the limit $\alpha\rightarrow0$,
which implies $M\rightarrow\infty$, $u\rightarrow0$, $\rho\rightarrow\pi$,
in the $(-)$ case $\xi\rightarrow\pi$, $R_{P2,e}^{(-)}\rightarrow R_{P1}^{(-)}$,
and in the $(+)$ case $\xi\rightarrow0$, $R_{P2,e}^{(+)}\rightarrow R_{P1}^{(+)}$.
The same result can be obtained by solving the Yang-Baxter equations.
$S_{P}$ is also bijective in the physical strip. The arguments used
to establish the results described in section \ref{sec 4.1.4} can
be applied to this case as well, and it is easy to see that the multiplicity
of a level $n$ boundary bound state is $2^{n}$ irrespectively of
the particular boundary fusing rules. The reflection amplitudes on
the boundary bound states are also free from physical strip poles
for any values of the fusing angles. However, if some value of the
boundary fusing angles occurs several times, then the statistical
properties of the particles should be taken into consideration. In
particular, if all the boundary fusing angles are the same (see \cite{key-6}),
then the multiplicity of a level $n$ boundary bound state is $2$
for each value of $n$. This is consistent with the multiplicities
obtained by taking the zero bulk coupling limit of the boundary sinh-Gordon
model.

\subsection{Boundary $a_{2}^{(1)}$ affine Toda field theory}

The bulk spectrum of this model contains two particles 1 and 2 of
equal mass $m_{1}=m_{2}$. Their fusing rules are $1+1\rightarrow2$
$(\pi/3)$ and $2+2\rightarrow1$ $(\pi/3)$. The charge conjugate
of 1 is 2. These rules are easily seen to be consistent with associating
supersymmetric particle representations to 1 and 2 with $\alpha=-1/(2M)$,
$m_{1}=m_{2}=2M\sin(\pi/3)$ \cite{key-1}. $M$ is a parameter of
the model.

The boundary version of this model described in \cite{key-7} contains
the boundary states $b_{n,m}$ for all $n,m\in\ZZ$, $n+m\geq0$,
$-\frac{1}{2B}-\hlf<n,m<\frac{1}{2B}+\hlf$, and the states $b_{-n,n}$
and $b_{n,-n}$ for all $n\in\ZZ$, $0\leq\frac{3}{2B}+\hlf$, where
$B$ is a parameter of the bulk model. We consider only generic values
of $B$ and only the domain $0<B<1$, as in \cite{key-7}. $b_{0,0}$
is the ground state. The fusing rules are shown in the following table:
\vspace{0.5cm}

\begin{center}\begin{tabular}{|c|c|c|c|}
\hline 
Initial state&
 particle&
 rapidity&
 final state\tabularnewline
\hline
$b_{n,m}$&
 1&
 $i\frac{\pi}{6}-i\frac{\pi}{6}B(2n+1)$&
 $b_{n+1,m}$\tabularnewline
\hline
$b_{n,m}$&
 2&
 $i\frac{\pi}{6}-i\frac{\pi}{6}B(2m+1)$&
 $b_{n,m+1}$\tabularnewline
\hline
$b_{-n,n}$, $0\leq n$&
 1&
 $i\frac{\pi}{2}-i\frac{\pi}{6}B(2n+1)$&
 $b_{-n-1,n+1}$\tabularnewline
\hline
$b_{n,-n}$, $0\leq n$&
 2&
 $i\frac{\pi}{2}-i\frac{\pi}{6}B(2n+1)$&
 $b_{n+1,-n-1}$ \tabularnewline
\hline
\end{tabular}\end{center}

\vspace{0.5cm}

The supersymmetric parts associated to these boundary bound states
are denoted by $Sb_{n,m}$, $Sb_{-n,n}$, $Sb_{n,-n}$ respectively.
It is straightforward to verify that the fusing rules above imply
that the corresponding sets $L_{Sb_{n,m}}$, $L_{Sb_{-n,n}}$, $L_{Sb_{n,-n}}$
are the following:

$L_{Sb_{n,m}}=\{ i(\frac{\pi}{3}-\frac{\pi}{6}B(2j+1)),i(\frac{\pi}{6}B(2j+1)),i(\frac{\pi}{3}-\frac{\pi}{6}B(2l+1)),i(\frac{\pi}{6}B(2l+1))\,|\, j=1...n,l=1...m\}$
if $n,m\geq0$,

$L_{Sb_{-n,n}}=L_{Sb_{n,-n}}=\{ i(\frac{2\pi}{3}-\frac{\pi}{6}B(2j+1)),i(\frac{\pi}{3}-\frac{\pi}{6}B(2j+1))\,|\, j=1...n\}$,

$L_{Sb_{n,m}}=\{ i(\frac{2\pi}{3}-\frac{\pi}{6}B(2j+1)),i(\frac{\pi}{3}-\frac{\pi}{6}B(2l+1)),i(\frac{\pi}{6}B(2r+1))\,|\, j=1...(-n),l=1...m,r=(-n+1)...m\}$
if $n<0$,

$L_{Sb_{n,m}}=\{ i(\frac{2\pi}{3}-\frac{\pi}{6}B(2j+1)),i(\frac{\pi}{3}-\frac{\pi}{6}B(2l+1)),i(\frac{\pi}{6}B(2r+1))\,|\, j=1...(-m),l=1...n,r=(-m+1)...n\}$
if $m<0$.

Consequently, the multiplicity of $Sb_{n,m}$ is $2^{n+m}$ if $n,m\geq0$,
$2^{m}$ if $n<0$, $2^{n}$ if $m<0$, provided that there is no
coincidence between $\pi-\xi$ and the rapidities above. $\pi-\rho_{max}/2=5\pi/6$,
so condition (\ref{feltetel 4}) is satisfied and the supersymmetric
factors of the reflection amplitudes have no poles on the imaginary
axis in the physical strip.

\subsection{Boundary $a_{4}^{(1)}$ affine Toda field theory}

The bulk spectrum of this model contains four particles 1, 2, 3, 4
of mass $m_{1}=m_{4}=2M\sin(\pi/5)$, $m_{2}=m_{3}=2M\sin(2\pi/5)$,
where $M$ is a parameter of the model. The fusing rules are $a+b\rightarrow c$,
where either $c=a+b$ or $c=a+b-5$. The corresponding fusing angles
are $\frac{\pi}{5}(a+b)$ if $c=a+b$ and $\frac{\pi}{5}(10-a-b)$
if $c=a+b-5$. The charge conjugate of 1 and 2 is 4 and 3. Theses
rules are easily seen to be consistent with associating supersymmetric
particle representations to 1, 2, 3, 4 with $\alpha=-1/(2M)$ \cite{key-1}.

The boundary version of this model described in \cite{key-7} has
two classes of inequivalent solitonic boundary conditions to which
different boundary spectra belong.

If the boundary conditions are of the first class, then there are
the boundary states $b_{n_{2},n_{3}}$, $n_{2},n_{3}\in\ZZ$, $n_{2}+n_{3}\geq0$,
$-\frac{1}{2B}-\hlf<n_{2},n_{3}<\frac{1}{2B}+\hlf$, and $b_{n,-n}$
and $b_{-n,n}$ for all $n\in\ZZ$, $0\leq n<\frac{5}{2B}+\hlf$,
where $B$ is a parameter of the bulk model. Generic values of $B$
are considered in the domain $0<B<1$. The fusing rules, that are
analogous to those of the $a_{2}^{(1)}$ model are listed in the following
table:\vspace{0.5cm}

\begin{center}\begin{tabular}{|c|c|c|c|}
\hline 
Initial state&
 particle&
 rapidity&
 final state\tabularnewline
\hline
$b_{n_{2},n_{3}}$&
 2&
 $i\frac{\pi}{10}-i\frac{\pi}{10}B(2n_{2}+1)$&
 $b_{n_{2}+1,n_{3}}$\tabularnewline
\hline
$b_{n_{2},n_{3}}$&
 3&
 $i\frac{\pi}{10}-i\frac{\pi}{10}B(2n_{3}+1)$&
 $b_{n_{2},n_{3}+1}$\tabularnewline
\hline
$b_{-n,n}$, $0\leq n$&
 1&
 $i\frac{\pi}{2}-i\frac{\pi}{10}B(2n+1)$&
 $b_{-n-1,n+1}$\tabularnewline
\hline
$b_{n,-n}$, $0\leq n$&
 4&
 $i\frac{\pi}{2}-i\frac{\pi}{10}B(2n+1)$&
 $b_{n+1,-n-1}$ \tabularnewline
\hline
\end{tabular}\end{center}

\vspace{0.5cm}

The supersymmetric parts associated to these particles are denoted
by $Sb_{n_{2},n_{3}}$, $Sb_{-n,n}$, $Sb_{n,-n}$ respectively. We
have

$L_{Sb_{n_{2},n_{3}}}=\{ i(\frac{\pi}{5}-\frac{\pi}{10}B(2j+1)),i(\frac{\pi}{10}B(2j+1)),i(\frac{\pi}{5}-\frac{\pi}{10}B(2l+1)),i(\frac{\pi}{10}B(2l+1))\,|\, j=1...n_{2},l=1...n_{3}\}$
if $n_{2},n_{3}\geq0$,

$L_{Sb_{-n,n}}=L_{Sb_{n,-n}}=\{ i(\frac{4\pi}{5}-\frac{\pi}{10}B(2j+1)),i(\frac{\pi}{5}-\frac{\pi}{10}B(2j+1))\,|\, j=1...n\}$,

$L_{Sb_{n_{2},n_{3}}}=\{ i(\frac{4\pi}{5}-\frac{\pi}{10}B(2j+1)),i(\frac{\pi}{5}-\frac{\pi}{10}B(2l+1)),i(\frac{\pi}{10}B(2r+1))\,|\, j=1...(-n_{2}),l=1...n_{3},r=(-n_{2}+1)...n_{3}\}$
if $n_{2}<0$,

$L_{Sb_{n_{2},n_{3}}}=\{ i(\frac{4\pi}{5}-\frac{\pi}{10}B(2j+1)),i(\frac{\pi}{5}-\frac{\pi}{10}B(2l+1)),i(\frac{\pi}{10}B(2r+1))\,|\, j=1...(-n_{3}),l=1...n_{2},r=(-n_{3}+1)...n_{2}\}$
if $n_{3}<0$.

The multiplicity of $Sb_{n_{2},n_{3}}$ is $2^{n_{2}+n_{3}}$ if $n_{2},n_{3}\geq0$,
$2^{n_{3}}$ if $n_{2}<0$, $2^{n_{2}}$ if $n_{3}<0$, provided that
there is no coincidence between $i(\pi-\xi)$ and the rapidities above.

If $B$ is sufficiently small, then the states $Sb_{-1,1},Sb_{-2,2},...$
violate condition (\ref{feltetel 4}), so the supersymmetric factor
of the reflection amplitudes of particle 1 on these states have poles
on the imaginary axis in the physical strip. There is one such pole
for $Sb_{-1,1}$ which is canceled by a zero of the non-supersymmetric
factor of the reflection amplitude, but there are two such poles for
$Sb_{-2,2}$, of which only one is canceled. Applying lemma 1 in \cite{key-20}
it is not hard to see that the remaining pole cannot be explained
by a Coleman-Thun mechanism. The SUSY factor of the reflection amplitude
of 1 on $Sb_{-n,n}$ has in general $n$ poles located at $i(\frac{\pi}{10}-\frac{\pi}{10}B(2m+1))$,
$m=1...n$ of which only the one at $i(\frac{\pi}{10}-\frac{3\pi}{10}B)$
is canceled by the non-supersymmetric factor, and they cannot be explained
by Coleman-Thun mechanism. The situation is similar if the ground
state reflection factor is $R_{P2,e}^{(+)}$.

If the boundary conditions are of the second class, then there are
the boundary states $b_{n_{1},n_{2},n_{3},n_{4}}$, $n_{1},n_{2},n_{3},n_{4}\in\ZZ$,
$n_{1}+n_{2}\geq0$, $n_{2}+n_{3}\geq0$, $n_{3}+n_{4}\geq0$. The
fusing rules are the following: \vspace{0.5cm}

\begin{center}\begin{tabular}{|c|c|c|c|}
\hline 
Initial state&
 particle&
 rapidity&
 final state\tabularnewline
\hline
$b_{n_{1},n_{2},n_{3},n_{4}}$&
 1&
 $i\frac{\pi}{10}-i\frac{\pi}{10}B(2n_{1}+1)$&
 $b_{n_{1}+1,n_{2},n_{3},n_{4}}$\tabularnewline
\hline
$b_{n_{1},n_{2},n_{3},n_{4}}$&
 2&
 $i\frac{\pi}{10}-i\frac{\pi}{10}B(2n_{2}+1)$&
 $b_{n_{1},n_{2}+1,n_{3},n_{4}}$\tabularnewline
\hline
$b_{n_{1},n_{2},n_{3},n_{4}}$&
 3&
 $i\frac{\pi}{10}-i\frac{\pi}{10}B(2n_{3}+1)$&
 $b_{n_{1},n_{2},n_{3}+1,n_{4}}$\tabularnewline
\hline
$b_{n_{1},n_{2},n_{3},n_{4}}$&
 4&
 $i\frac{\pi}{10}-i\frac{\pi}{10}B(2n_{4}+1)$&
 $b_{n_{1},n_{2},n_{3},n_{4}+1}$\tabularnewline
\hline
$b_{-n_{2},n_{2},n_{3},n_{4}}$, $0\leq n_{2}$&
 1&
 $i\frac{3\pi}{10}-i\frac{\pi}{10}B(2n_{2}+1)$&
 $b_{-n_{2}-1,n_{2}+1,n_{3}n_{4}}$\tabularnewline
\hline
$b_{n_{1},n_{2},n_{3},-n_{3}}$, $0\leq n_{3}$&
 4&
 $i\frac{3\pi}{10}-i\frac{\pi}{10}B(2n_{3}+1)$&
 $b_{n_{1},n_{2},n_{3}+1,n_{3}-1}$ \tabularnewline
\hline
\end{tabular}\end{center}

\vspace{0.5cm} Using our rules we get

$L_{Sb_{n_{1},n_{2},n_{3},n_{4}}}=\{ i(\frac{2\pi}{5}-\frac{\pi}{10}B(2l_{1}+1)),i(\frac{\pi}{5}+\frac{\pi}{10}B(2l_{1}+1)),i(\frac{2\pi}{5}-\frac{\pi}{10}B(2l_{4}+1)),i(\frac{\pi}{5}+\frac{\pi}{10}B(2l_{4}+1)),$

$i(\frac{\pi}{5}-\frac{\pi}{10}B(2l_{2}+1)),i(\frac{\pi}{10}B(2l_{2}+1)),i(\frac{\pi}{5}-\frac{\pi}{10}B(2l_{3}+1)),i(\frac{\pi}{10}B(2l_{3}+1))\,|\, l_{i}=1...n_{i}\}$
if $n_{1},n_{2},n_{3},n_{4}\geq0$,

$L_{Sb_{-n_{2},n_{2},n_{3},n_{4}}}=\{ i(\frac{\pi}{5}-\frac{\pi}{10}B(2l_{3}+1)),i(\frac{\pi}{10}B(2l_{3}+1)),i(\frac{2\pi}{5}-\frac{\pi}{10}B(2l_{4}+1)),i(\frac{\pi}{5}+\frac{\pi}{10}B(2l_{4}+1)),$

$i(\frac{6\pi}{10}-\frac{\pi}{10}B(2l_{2}+1)),i(\frac{\pi}{10}B(2l_{2}+1))\,|\, l_{3}=1...n_{3},l_{4}=1...n_{4},l_{2}=1...n_{2}\}$
if $n_{3},n_{4}\geq0$,

$L_{Sb_{n_{1},n_{2},n_{3},-n_{3}}}$ is similar to $L_{Sb_{-n_{2},n_{2},n_{3},n_{4}}}$,

$L_{Sb_{-n_{2},n_{2},n_{3},-n_{3}}}=\{ i(\frac{6\pi}{10}-\frac{\pi}{10}B(2l_{2}+1)),i(\frac{\pi}{10}B(2l_{2}+1)),i(\frac{6\pi}{10}-\frac{\pi}{10}B(2l_{3}+1)),i(\frac{\pi}{10}B(2l_{3}+1))\,|\, l_{2}=1...n_{2},l_{3}=1...n_{3}\}$,

$L_{Sb_{n_{1},n_{2},n_{3},n_{4}}}=\{ i(\frac{\pi}{5}-\frac{\pi}{10}B(2l_{3}+1)),i(\frac{\pi}{10}B(2l_{3}+1)),i(\frac{2\pi}{5}-\frac{\pi}{10}B(2l_{4}+1)),i(\frac{\pi}{5}+\frac{\pi}{10}B(2l_{4}+1)),$

$i(\frac{6\pi}{10}-\frac{\pi}{10}B(2j_{1}+1)),i(\frac{\pi}{10}B(2j_{2}+1)),i(\frac{\pi}{5}+\frac{\pi}{10}B(2j_{3}+1))\,|\, l_{3}=1...n_{3},l_{4}=1...n_{4},j_{2}=1...n_{2},j_{1}=1...(-n_{1}),j_{3}=(-n_{1}+1)...n_{2}\}$
if $n_{3},n_{4},n_{2}\geq0$, $n_{1}<0$,

the remaining two cases when $n_{4}<0$, $n_{1}\geq0$ and $n_{4},n_{1}<0$
are similar.

The multiplicity of $Sb_{n_{1},n_{2},n_{3},n_{4}}$ is $2^{n_{1}+n_{2}+n_{3}+n_{4}}$
if $n_{1},n_{2},n_{3},n_{4}\geq0$, $2^{n_{2}+n_{3}+n_{4}}$ if $n_{1}<0$,
$n_{4}\geq0$, $2^{n_{1}+n_{2}+n_{3}}$ if $n_{1}\geq0$, $n_{4}<0$
and $2^{n_{2}+n_{3}}$ if $n_{1},n_{4}<0$ provided that there is
no coincidence between $\pi-\xi$ and the rapidities above. Condition
(\ref{feltetel 4}) is satisfied in this case, so the supersymmetric
factors of the reflection amplitudes have no poles on the imaginary
axis in the physical strip.

\section{Discussion }

We considered the boundary supersymmetric bootstrap programme in the
case when the ground state is a singlet with RSOS label $\hlf$ and
the bulk particles transform in the kink and boson/fermion representations.

We presented a review of the relevant analogous results for bulk bootstrap.
We introduced the boundary supersymmetry algebra and its action in
the framework proposed by \cite{key-13,key-14}, which requires that
the boundary supersymmetry algebra has to be a coideal of the bulk
SUSY algebra. In accordance with the literature, we found that there
are two possible boundary supersymmetry algebras. The corresponding
two cases ---denoted by $(+)$ and $(-)$--- lead to different supersymmetric
ground state reflection factors. We found that these factors are essentially
the same as those given in \cite{key-10,key-11,key-5}. Although the
algebras of the $(+)$ and $(-)$ cases appear to play symmetric role,
the corresponding kink reflection factors turn out to be significantly
different. A further important difference between the two cases is
that in the $(+)$ case the boson/fermion reflection factor can be
obtained by bootstrap from the kink reflection factor only at special
values of its parameters \cite{key-11}. We also found that the kink
and boson/fermion reflection factors can be degenerate at particular
rapidities depending on a parameter $\gamma$ of the ground state
representation.

We presented supersymmetric boundary fusing rules by which the representations
and reflection factors for excited boundary bound states can be easily
determined in specific models. The main difficulty of the problem
is to handle the degeneracies of the boundary fusing tensors that
occur at particular rapidities (resulting from the degeneracies of
the one-particle reflection factors). These degeneracies are closely
related to the degeneracies of the bulk two-particle scattering factors
and of the ground state one-particle reflection factors. We found
that the boundary fusing rules are analogous to the bulk rules \cite{key-1,key-3},
and that it is useful to characterize the boson/fermion multiplets
by their constituent kinks.

We considered the simplest case, when the two-particle scattering
factors and ground state reflection factors are minimal and have no
poles and overall zeroes on the imaginary axis in the physical strip,
and there is no interplay between the poles and zeroes of the supersymmetric
and non-supersymmetric factors of the S-matrix and reflection amplitude.
We found that the main restriction on the applicability of the described
construction follows from this condition and from the condition that
the ground state is a singlet with RSOS label $\hlf$ (whereas in
the bulk the main restriction arises because the supersymmetry representations
are prescribed).

We applied our results to the boundary sine-Gordon model \cite{key-16,key-2},
to the $a_{2}^{(1)}$ and $a_{4}^{(1)}$ affine Toda field theories
\cite{key-7} and to the free particle. We found that the boundary
$a_{2}^{(1)}$ affine Toda model admits a tensor product type supersymmetrization,
whereas the minimal supersymmetrization of the $a_{4}^{(1)}$ model
with first class boundary condition is not possible: the supersymmetric
reflection factors on some excited boundary states introduce poles
that cannot be explained by Coleman-Thun mechanism. It is an open
problem whether the supersymmetrized $a_{2}^{(1)}$ and $a_{4}^{(1)}$
reflection amplitudes describe any Lagrangian field theory. We also
considered the sinh-Gordon model briefly.

It is likely that representations beyond the kink and boson/fermion
for the bulk particles and other (possibly non-singlet) ground state
representations can also be relevant to some models.

\subsection*{Acknowledgments}

The author would like to thank L. Palla, G. Tak\'{a}cs and Z. Bajnok
for proposing this problem, for very useful discussions, and for comments
on the manuscript.

\section*{Appendix}

\[
G^{[i,j]}(\theta)=R^{[i,j]}(\theta)R^{[i,j]}(\pi i-\theta)\]
 \[
R^{[i,j]}(\theta)=\frac{1}{\Gamma(\frac{\theta}{2\pi i})\Gamma(\frac{\theta}{2\pi i}+\hlf)}\prod_{k=1}^{\infty}\frac{\Gamma(\frac{\theta}{2\pi i}+\Delta_{1}+k+1)\Gamma(\frac{\theta}{2\pi i}-\Delta_{1}+k)}{\Gamma(\frac{\theta}{2\pi i}+\Delta_{1}+k-\hlf)\Gamma(\frac{\theta}{2\pi i}-\Delta_{1}+k+\hlf)}\times\]
 \[
\times\frac{\Gamma(\frac{\theta}{2\pi i}+\Delta_{2}+k-\hlf)\Gamma(\frac{\theta}{2\pi i}-\Delta_{2}+k-\hlf)}{\Gamma(\frac{\theta}{2\pi i}+\Delta_{2}+k)\Gamma(\frac{\theta}{2\pi i}-\Delta_{2}+k)},\]
 where \[
\Delta_{1}=(u_{i}+u_{j})/(2\pi),\qquad\Delta_{2}=(u_{i}-u_{j})/(2\pi).\]

\[
ZX^{(+)}(\theta)=\sqrt{m}P(\theta+i\rho/2)P(\theta-i\rho/2)\sqrt{2}K(2\theta)2^{-\theta/(i\pi)},\]
 where $m=2M\cos(\frac{\rho}{2})$, $0\leq\rho<\pi$, $M=-1/\alpha$.

\[
\tilde{Z}^{(-)}(\theta)=K(2\theta)2^{-\theta/(i\pi)}F(\theta-i\rho/2)F(\theta+i\rho/2),\qquad F(\theta)=P(\theta)K(\theta+i\xi)K(\theta-i\xi),\]
 where $\gamma=-2\sqrt{M}\cos\frac{\xi}{2}$, $0\leq\xi\leq\pi$ or
$Re(\xi)=0$ or $Re(\xi)=\pi$.\[
ZX^{(-)}(\theta)=\sqrt{m}\tilde{Z}^{(-)}(\theta)\frac{1}{\sqrt{2}}(\cos(\rho/2)-\cosh(\theta)),\]
 where $\xi=\pi$. \[
\tilde{Z}^{(+)}(\theta)=iK(2\theta)2^{-\theta/(i\pi)}F(\theta-i\rho/2)F(\theta+i\rho/2)U(\theta),\qquad F(\theta)=P(\theta)K(\theta+i\xi)K(\theta-i\xi),\]
 where $\gamma=-2\sqrt{M}i\sin(\xi/2)$, $-\pi\leq\xi\leq\pi$ or
$Re(\xi)=0$ or $Re(\xi)=\pm\pi$, \[
U(\theta)=f(\theta)/f(-\theta),\]
 \[
\frac{U(i\theta)}{U(i\theta-i\pi)}=-\frac{\cos(\theta)-\cos(\rho/2)}{\cos(\theta)+\cos(\rho/2)},\]
 \[
f(\theta)=\prod_{k=1}^{\infty}\frac{\Gamma(\frac{\rho}{4\pi}-\frac{\theta}{2\pi i}-\frac{1}{2}+k)\Gamma(\frac{\rho}{4\pi}+\frac{\theta}{2\pi i}+k)\Gamma(-\frac{\rho}{4\pi}-\frac{\theta}{2\pi i}-\frac{1}{2}+k)\Gamma(-\frac{\rho}{4\pi}+\frac{\theta}{2\pi i}+k)}{\Gamma(\frac{\rho}{4\pi}-\frac{\theta}{2\pi i}+k)\Gamma(\frac{\rho}{4\pi}+\frac{\theta}{2\pi i}+\frac{1}{2}+k)\Gamma(-\frac{\rho}{4\pi}-\frac{\theta}{2\pi i}+k)\Gamma(-\frac{\rho}{4\pi}+\frac{\theta}{2\pi i}+\frac{1}{2}+k)}.\]

\[
K(\theta)=\frac{1}{\sqrt{\pi}}\prod_{k=1}^{\infty}\frac{\Gamma(k-\frac{1}{2}+\frac{\theta}{2\pi i})\Gamma(k-\frac{\theta}{2\pi i})}{\Gamma(k+\frac{1}{2}-\frac{\theta}{2\pi i})\Gamma(k+\frac{\theta}{2\pi i})},\]
 \[
P(\theta)=\prod_{k=1}^{\infty}\left[\frac{\Gamma(k-\frac{\theta}{2\pi i})^{2}}{\Gamma(k-\frac{1}{4}-\frac{\theta}{2\pi i})\Gamma(k+\frac{1}{4}-\frac{\theta}{2\pi i})}/\{\theta\leftrightarrow-\theta\}\right].\]

\end{document}